\documentclass[aps,showpacs,twocolumn]{revtex4}

\usepackage{bm}
\usepackage{amsmath}
\usepackage{amssymb}
\usepackage{graphicx}
\usepackage{epsfig}
\usepackage{epstopdf}
\begin{document}

\def\pr{\prime}
\def\be{\begin{equation}}
\def\en#1{\label{#1}\end{equation}}
\def\d{\dagger}
\def\bar#1{\overline #1}
\newcommand{\per}{\mathrm{per}}

\newcommand{\rd}{\mathrm{d}}
\newcommand{\vare}{\varepsilon }

\title{ Partial indistinguishability theory for   multi-photon  experiments in multiport  devices    }

\author{V. S. Shchesnovich}

\address{Centro de Ci\^encias Naturais e Humanas, Universidade Federal do
ABC, Santo Andr\'e,  SP, 09210-170 Brazil }

\begin{abstract} 
 We  generalize  an approach   for  description of    multi-photon   experiments  with  multi-port unitary linear optical devices, started in \textit{Phys. Rev. A \textbf{89}, 022333 (2014)}   with   single  photons  in mixed spectral states, to arbitrary (multi-photon) input and arbitrary photon detectors.      We show that output probabilities are  \textit{always}  given  in terms of the matrix permanents of the Hadamard product  of a matrix built from the  network  matrix  and  matrices  built from     spectral state of   photons  and   spectral sensitivities of  detectors.   Moreover, in case of input with up to one photon per mode, the output probabilities  are given by  a sum (or integral) with  each term being   the absolute value squared of  such a   matrix permanent.  We   conjecture that, for an arbitrary multi-photon input,  zero output probability of an output configuration  is  \textit{always} the result of an exact cancellation of  quantum transition   amplitudes   of   completely  indistinguishable photons  (a subset of all input photons) and, moreover,   \textit{ is   independent}  of   coherence between  only partially indistinguishable photons. The conjecture is supported by examples. Furthermore, we propose a measure of partial indistinguishability of photons which generalizes  Mandel's observation,  and   find the law of degradation of  quantum coherence  in  a  realistic Boson-Sampling device     with increase of  the total number of photons and/or their  ``classicality parameter".

 \end{abstract}

\maketitle

\section{Introduction}
\label{sec1}

It is well-known  \cite{Mandel1991} that  quantum coherence of an electromagnetic field and  indistinguishability of  photons are intimately related to each other. The most famous quantum coherence effect of this type is  the Hong-Ou-Mandel (HOM) dip \cite{HOM,Mandel1999}, where the ``dip" in the  output  coincidence probability of a balanced beam-splitter  corresponds to complete indistinguishability of  single photons at its input. Many important developments  in the  area of multi-photon experiments with multi-port optical devices have been  achieved since then.   A generalization of the HOM effect and  a difference in  behavior of bosons and fermions was  analyzed    for   Bell multiport beam splitters  \cite{Mult1,Mult2,LB}.  An approach describing  partial distinguishability of   photons  obtained from the parametric down conversion sources was developed in Refs. \cite{Ou2006,Ou2008}.  Recently, a   zero transmission law due to a symmetry of the network matrix  \cite{ZTL} and  a quantum suppression law in many-particle interferences beyond the  boson and fermion statistics   were found \cite{MPI}.   Recent  advances in   quantum interference experiments   in linear multiport devices include   characterizing  temporal distinguishability of four and six photon states \cite{Xiang2006}, experimental control over eight individual single photons \cite{Yao2012},  observation of the two-photon HOM effect  on integrated 3 and 4 port devices \cite{Meany2012},  verification of the three-photon HOM effect  and the zero transmission law   on a tritter \cite{Spangolo2013},   three-photon quantum interference experiment  on an integrated eight-mode optical device \cite{Metcalf2013}, and observation of detection-dependent multi-photon coherence times \cite{Ra2013}.    The multi-photon quantum interference is  central in  the Boson-Sampling computer \cite{AA} with indistinguishable single photons and linear optics, the output of which is hard to simulate on a classical computer.  Recently the  experimental realization of the Boson-Sampling computer  was tested  on a small scale  \cite{E1,E2,E3,E4,E5}.   One must also mention the  well-known   proposal of the universal quantum computation with linear  optics \cite{KLM}.  

The above described      advances with  multi-photon experiments  of  increasing  complexity  (see also  the review \cite{MultPhExp})  and also the  recent achievements  in   fabrication of photon  sources \cite{SPS}  necessitate a theoretical approach which enables one to  account for    the effect of partial indistinguishability of  photons in a realistic general setup of a  multi-port  device with  an arbitrary   multi-photon input and  with account for  imperfect detectors. Here such a general  approach is developed by generalization of that of Ref. \cite{NDBS}.   As in  Refs. \cite{Ou2006,Ou2008,NDBS,SU3,SU3A} we  employ the permutation symmetry of  spectral state of photons to characterize their partial   indistinguishability  and  further advance this relation: we  derive the general output probabilities for multi-photon experiments with multi-port devices for an arbitrary number of network modes and an arbitrary  multi-photon input,  study the physical meaning of the partial indistinguishability  matrix, first introduced in  Ref. \cite{NDBS}, and introduce an auxiliary Hilbert space representation of  spectral states   of  photons, which  allows one to rewrite   output probabilities in a clear compact  form.   In view of   application to the Boson-Sampling experiments, we discuss in detail   the case of  input with at most one  photon per mode, give output probability in a simplified form, and study  degradation of quantum interference on a classicality parameter and  the total number of  photons.   Note that a different  approach  based on  the orthonormalization  of photon spectral states, used  in Refs. \cite{Ra2013,Add1, Add2}, which is  helpful in few-photon cases,  does not have a clear physical interpretation  and will not be of much help  for larger $N$  or   mixed spectral states  \cite{Tichy2014}.  

Since the symmetric (i.e., permutation) group  is the key object in our approach, one might expect   that usage of  advanced   features of the symmetric group (i.e., the group  characters  and the corresponding  Young diagrams)  is essential for understanding   multi-photon experiments in multi-port devices. Indeed,  recently  three-photon interference in a tritter was analyzed using  some    advanced symmetric  group  structures called  the matrix immanants  (related to the nontrivial group characters) \cite{SU3,SU3A}. However, such an approach is not scalable, since   Young diagrams   associated  with nontrivial group characters can only be  analyzed case by case with no formula for the general solution.   Our approach, on the other hand, \textit{does not depend} on  any of such advanced group structures. Only some elementary facts about the permutation group, such as the cycle decomposition,  are used.     We  show, for instance,  that the zero coincidence condition for  partially indistinguishable photons   of Ref.~\cite{SU3,SU3A}, involving  the matrix immanants,   can be   restated as a  zero permanent condition of a Hadamard product of network  matrix and a matrix built from   spectral states of  photons and  detector sensitivities.  
We  also  conjecture that  zero output probability of an output configuration  is  \textit{always} the result of exact cancellation of  quantum transition   amplitudes   of   completely  indistinguishable photons  (a subset of all input photons)  when  a network allows for such an exact cancellation. Moreover, in all cases,    zero probability   \textit{ is independent}  on degree of  coherence  of   only partially indistinguishable photons.
 
Finally, it should be mentioned that   the effect of partial indistinguishability of photons on  probabilities  at a network output  has a deep relation with the    duality (complementarity) between the fringe visibility and the which-way information.  This duality  is well-understood for  two-path interference experiments~\cite{GY1988,JSV1995,Englert1996}.    Indeed,   though the output probability is related to  a Glauber's  higher-order coherence function \cite{Glauber}, whereas the duality pertains to the first-order coherence of a single  quantum object, when all photons are detected for an input with a certain number of photons, one can reinterpret the   multi-photon interference as a multi-path interference experiment, where there are  $N!$ paths   for $N$ photons.   Such a  relation was studied  by Mandel \cite{Mandel1991} for $N=2$ (see also Ref. \cite{Loudon}).   However, following this   point of view   in discussion of  $N$-photon multi-port experiments for $N>2$ meets with several obstacles and is not pursued here.  One of them is that  generalization of the    duality  to multi-path   coherence is not   unique \cite{Englert2008}.    However, the duality supplies a clear physical interpretation of the formulas derived below. Moreover,  an  argument referring to the duality is used for formulation of the zero probability conjecture. 

In section \ref{sec2} we derive the general formula for output probability in a multi-mode network for arbitrary multi-photon input. Some details of the derivation are placed in appendix \ref{appA}.  In subsections \ref{sec2A} and \ref{sec2B} we  compare  the case of ideal (i.e., maximally efficient) detectors with that of  realistic detectors  for  two extreme cases of  input: the completely indistinguishable photons and  the maximally distinguishable photons. In subsection \ref{secPER} we express  output probability  via matrix permanents of the Hadamard product of  matrices, one built from network matrix and another  from spectral states of photons and sensitivities of detectors. In subsection \ref{sec2C} we propose a measure for partial indistinguishability of photons generalizing Mandel's parameter for $N>2$ photons.  
We focus on the input with a single photon or vacuum per input mode in section \ref{sec3}, where we give  a simplified   formula for   output probability and analyze its structure  for single photons in pure spectral states, subsection  \ref{sec3A}, and generalize the result to the case of single photons in mixed spectral states, subsection \ref{sec3B}.    In subsection \ref{secZero} we formulate zero probability conjecture and study few examples supporting it.   Some mathematical  calculations of section \ref{sec3} are relegated to appendices \ref{appB} and \ref{appC}. Finally, in subsection \ref{sec3C} we discuss   a model of the Boson-Sampling computer and compute  purity of the partial indistinguishability matrix as a measure of closeness  of a realistic device  with only partially indistinguishable photons to the ideal Boson-Sampling computer.  Some final remarks     are placed  in the concluding section \ref{con}.

\section{Output probability formula for fixed number of photons in each  input mode }
\label{sec2}

\textit{Input state.} -- Consider   a linear unitary  optical network of $M$ different  inputs (we  consider each input  to be  single mode)  where a $n_k$-photon state is injected into the $k$th input mode. Below we set $n_1+\ldots +n_M  = N$  (in general, the number of modes with a  nonvacuum input  is  less than $M$).  We are interested in the expression for  output probabilities for  such an input. In view of the problem formulation, it is convenient to use a basis for photon states consisting of  spatial mode $k$, polarization state $s$ (where, say,  $s=0$ and $s=1$ correspond to two orthogonal basis states of photon  polarization) and  frequency 
$\omega$. Denote  photon  creation and annihilation operators   in this basis  by a subscript $(k,s)$  and consider them  to be functions of $\omega$. A spatial unitary network can be defined  by an  unitary transformation between  input $a^\dag_{k,s}(\omega)$ and  output $b^\dag_{k,s}(\omega)$ photon creation operators, we set $a^\dag_{k,s}(\omega) = \sum_{l=1}^M U_{kl} b^\dag_{l,s}(\omega)$, where $U_{kl}$ is the unitary matrix describing such an optical network. Below we will employ vector notations for greater convenience, e.g., $\vec{n} = (n_1,\ldots,n_M)$ for  numbers of photons in spatial modes, $\vec{\omega}=(\omega_1,\ldots, \omega_N)$ for   frequencies, and  $\vec{s} = (s_1,\ldots,s_N)$ for   polarizations. Define also  $|\vec{n}|\equiv n_1+\ldots+n_M$ and  $\mu(\vec{n}) \equiv   \prod_{k=1}^M n_k!$. The general  $N$-photon input (a mixed state) with a certain number of photons in each  input mode  is given by the following expression
\begin{eqnarray}
\label{E1}
\rho(\vec{n}) &=& \frac{1}{\mu(\vec{n})} \sum_{\vec{s}{\,}^\pr}\sum_{\vec{s}}\int\rd\vec{\omega}^\pr\int\rd\vec{\omega}G(\vec{s}^{\,\pr},\vec{\omega}^\pr|\vec{s},\vec{\omega})\nonumber\\
&&\times \left[\prod_{\alpha=1}^N a^\d_{k_\alpha,s^\pr_\alpha}(\omega_\alpha^\pr)\right]|0\rangle\langle 0| \left[\prod_{\alpha=1}^N a_{k_\alpha,s_\alpha}(\omega_\alpha) \right],\quad 
\end{eqnarray}
where $k_1,\ldots,k_N$ are input  modes (generally repeated where the repetition numbers are given by $\vec{n}$) and $G$ is  a function describing  spectral and polarization   state (mixed, in general) of $N$ input photons~\footnote{To use index $k$   instead  of the double index $k_\alpha$  would not be simpler   because we have multiple  $k$-indices  for $n_k>1$. To  leave just the subindex by definition     $\hat{U}_{\alpha,\beta}\equiv U_{k_\alpha,l_\beta}$ is also not convenient, since some  formulae  have   essential dependence on  the output configuration of spatial modes due to different detectors attached to them.}. An  immediate consequence of the bosonic commutation relations is that any permutation $\pi$ of frequencies and polarizations associated with either creation or annihilation operators in Eq. (\ref{E1}), i.e.,  $(s_\alpha,\omega_\alpha)  \to (s_{\sigma(\alpha)},\omega_{\sigma(\alpha)})$, which permutes  photons from the same input mode $k$,  leaves the function $G$  of Eq.~(\ref{E1}) invariant. The group of such permutations, a subgroup of all permutations $\mathcal{S}_N$,  is equivalent to the tensor product of groups $\mathcal{S}_{n_1} \otimes\ldots\otimes \mathcal{S}_{n_M}$ (some $\mathcal{S}_{n_\alpha}$ may be  empty  due to $n_\alpha=0$).
Given  this permutation symmetry of $G$, the following normalization condition can be derived from the  fact that $\rho$ of Eq.~(\ref{E1}) is a density matrix with trace equal to one
\be  
\sum_{\vec{s}} \int\rd \vec{\omega}\, G(\vec{s},\vec{\omega}|\vec{s},\vec{\omega}) = 1. 
\en{E2}
The function  $G$ is also  constrained by  positivity of  the associated   density matrix $\rho$. Below we will  frequently use two other representations of the density matrix in Eq. (\ref{E1}). The   diagonalized form    
\begin{eqnarray}
\label{E1A}
&&\rho(\vec{n}) = \sum_{i} p_i |\tilde{\Phi}_i \rangle\langle\tilde{\Phi}_i |,\quad \langle\tilde{\Phi}_i|\tilde{\Phi}_j\rangle=\delta_{ij},\nonumber\\
&& \!\! | \tilde{\Phi}_i\rangle = \frac{1}{\sqrt{\mu(\vec{n})}}\sum_{\vec{s}}\int\rd\vec{\omega} \, {\Phi}_i( \vec{s},\vec{\omega}) \left[\prod_{\alpha=1}^N a^\d_{k_\alpha,s_\alpha}(\omega_\alpha)\right]|0\rangle,\nonumber\\
&&
\end{eqnarray}
 obtained by  decomposing   function $G$ of Eq. (\ref{E1}) as follows   $G(\vec{s}{\,}^\pr,\vec{\omega}^\pr|\vec{s},\vec{\omega}) = \sum_i p_i {\Phi}_i(\vec{s}{\,}^\pr,\vec{\omega}^\pr)  {\Phi}^*_i(\vec{s},\vec{\omega})$, where   $\sum_{\vec{s}}\int\rd \vec{ \omega} |{\Phi}_i(\vec{s},\vec{\omega})|^2 =1$ and $\sum_i p_i =1$ 
($p_i>0$), and    another very important representation, which applies to sources with some fluctuating parameter(s), say $x$.  In the latter case, the density matrix has a form similar to that of Eq. (\ref{E1A})  but with  some non-orthogonal states  $|\tilde{\Phi}(x)\rangle$   
\be
\rho (\vec{n}) = \int\rd x\, p(x) |\tilde{\Phi}(x) \rangle\langle\tilde{\Phi}(x)|,
\en{E1B}
where we assume that the state vector $|\tilde{\Phi}(x)\rangle$ is given similar as in the second line of Eq. (\ref{E1A}).

Typical input states encountered in experiments are covered by the input of  Eq. (\ref{E1A}) or (\ref{E1B}). For instance,   if we have $N$ independent  sources of single photons  attached to modes $k_\alpha$, $\alpha=1,\ldots,N$,  with source $\alpha$  emitting single photons in a polarized (say $s_\alpha=1$) Gaussian state with the  central frequency $\Omega_\alpha$, spectral width $\Delta_\alpha$,   and arrival time $t_\alpha$, then the corresponding  input state is pure, $\rho = |\tilde{\Phi}\rangle\langle\tilde{\Phi}|$, where
\be
 |\tilde{\Phi}\rangle =\int\rd\vec{\omega}\left[\prod_{\alpha=1}^N  \phi_\alpha(\omega_\alpha) a^\d_{k_\alpha,1}(\omega_\alpha)\right]|0\rangle,
 \en{Ex1a}
with 
\be 
\phi_\alpha(\omega) = (2\pi\Delta_\alpha^2)^{-\frac14}\exp\left\{i\omega t_\alpha - \frac{(\omega-\Omega_\alpha)^2}{4\Delta^2_\alpha}\right\}
\en{Ex1b}
(note that we write $\phi_\alpha(\omega)$ and not $\phi_\alpha(\omega,t_\alpha)$  since it is a function of $\omega$,  whereas  $t_\alpha$ is a fixed parameter , different for different  index  $\alpha$; we will use this rule below for the sake of  simplicity). 
One frequent example of this kind is of  $N$  photons in   the same   Gaussian state, i.e.,  $\Omega_\alpha=  \Omega$ and $ \Delta_\alpha = \Delta$, but  
 with   different arrival times.   This example is,  of course, only illustrative and sometimes used to model a realistic situation due to manageability  of the Gaussian function and that in experiments only few parameters, such as the central frequency and  spectral width of the photon sources are known with some  precision. One can contemplate a more general model of this kind,  when    polarized single photons have the  spectral states  of the same shape, differing  only by the delay time, the appropriate  representation is  $\phi_\alpha(\omega) =  \int\rd t e^{i\omega t} f(t- t_\alpha)$   for an  arbitrary function $f(t)$ with the norm equal    to 1.  
 
When  spectral states of photons   have fluctuating parameters, e.g.,  the arrival time,  polarization, etc.,  the most appropriate representation is Eq. (\ref{E1B}).  For example, such an input gives a model of realistic Boson-Sampling computer \cite{AA} (see section \ref{sec3C} for more details).

\textit{Output probabilities and   interference of ``paths''.} -- Consider $M$, generally different,  number resolving detectors attached to  network output modes.  The probability  of detecting $m_1,...,m_M$ photons in  network output modes $1,\ldots,M$  can be derived using  the   photon counting  theory \cite{Glauber,Mandel,KK,VWBook}. The result is that  the probability for all  photons to be  detected at network  output in a configuration $\vec{m}$ is given  by the  expectation value of the following operator (see also  appendix  A in  Ref. \cite{NDBS})
\begin{eqnarray}
\label{E3}
&&\Pi(\vec{m}) = \frac{1}{\mu(\vec{m})}\sum_{\vec{s}}\int  \rd \vec{\omega} \prod_{\alpha=1}^N\Gamma_{l_\alpha}(s_\alpha,\omega_\alpha)\nonumber\\
&&\times\left[ \prod_{\alpha=1}^N b^\dag_{l_\alpha,s_\alpha}(\omega_\alpha)\right]|0\rangle\langle 0|\left[\prod_{\alpha=1}^N b_{l_\alpha,s_\alpha}(\omega_\alpha)\right],
\end{eqnarray}
where the indices  $l_1,...,l_N$    comprise   the sequence  $1,...,1,2,...,2,...,M,...,M$, with each index $j$ appearing $m_j$ times, and  $0\le \Gamma_l(s,\omega)\le 1$ is the   sensitivity function of the  detector attached to the $l$th output mode. The output probability of a configuration  $\vec{m}$ reads
\begin{eqnarray}
\label{E4}
&& P(\vec{m}|\vec{n}) =  \mathrm{Tr}\{ \rho(\vec{n}) \Pi(\vec{m})\}.
\end{eqnarray} 
 The  operator $\Pi(\vec{m})$ in Eq. (\ref{E4}) is  Hermitian and positive, but such operators  generally do not sum up to the identity operator (more precisely, to the projector on the symmetric subspace corresponding to  $N$ bosons).  However, for   efficient  detectors, when  all output  photons are detected,  each $\Pi(\vec{m})$ becomes    an element  of the positive operator valued measure  realizing  the above described detection.  In this case the probabilities in Eq. (\ref{E4}) sum to $1$ under the constraint $|\vec{m}| = N$.

The essence of our  approach below is based on the fact that  basis   variables $(k,s,\omega)$ are divided  into two parts: (i) spatial mode $k$, affected by  an unitary network, and    (ii) spectral part  (functions of polarization and frequency), not changed by the network and thus    serving as a  label for  partial indistinguishability of photons (by the distinguishability here and below  we mean    distinguishability  detectable in an  experiment    in the above described setting). The Fock space, natural for identical particles,  is not the most appropriate Hilbert space for treating partial indistinguishability, since it involves the boson creation and annihilation operators indexed by   $(k,s,\omega)$, whereas  only  spectral part defines     partial indistinguishability of photons. Another problem with the Fock space  is that to treat partial indistinguishability it is   better to employ a basis used for distinguishable particles. Below we employ  such an auxiliary Hilbert  space of $N$ fictitious distinguishable particles to use for description of  spectral state  of $N$ photons.  In this way a connection to the duality of the which-path  information vs. the interference  visibility can  be established: one can visualize  the   transitions    through a unitary network as ``paths" (there are $N!$ paths which can be labelled by elements of the symmetric group $\mathcal{S}_N$) whereas the  spectral states serve as some  internal degrees of freedom    which can, in principle,  be observed by  the environment. Summation over the path amplitudes  is affected by indistinguishability of  spectral states of photons  and also by spectral sensitivities of detectors.  For identical detectors, two permutations of the   fictitious particles, one  at input ($\sigma_1$) and one at output ($\sigma_2$),  represent a different set of paths with respect to $\sigma_1=\sigma_2=I$ (identity permutation) only if they are not equal (spectral data are not changed by the network).  But for different detectors even  if $\sigma_2 = \sigma_1$ the output probability is generally different for different  $\sigma_1$.   
Hence, a $N!\times N!$-dimensional partial indistinguishability matrix, indexed by elements of $\mathcal{S}_N$, describes all possible path interferences  for general detectors, whereas,  at most   $N!$ parameters of such a matrix are different  for identical detectors. 

Now let us give  output probability for an arbitrary input Eq. (\ref{E1}).  Due to the relation  $a^\dag_{k,s}(\omega) = \sum_{l=1}^M U_{kl} b^\dag_{l,s}(\omega)$ between   input and output modes, Eq. (\ref{E4}) is a nonnegative quadratic form with complex arguments  equal to  products of $N$   matrix elements of  a network matrix $U$, where the spectral part defines the matrix of this quadratic form. We have   from Eqs.~(\ref{E1}), (\ref{E3}), and (\ref{E4}) (the details can be found in appendix  \ref{appA})
\begin{eqnarray}
\label{E5}
P(\vec{m}|\vec{n}) &=& \frac{1}{\mu(\vec{m})\mu(\vec{n})}\sum_{\sigma_1}\sum_{\sigma_2} J(\sigma_1,\sigma_2) \nonumber\\
&& \times \prod_{\alpha=1}^NU^*_{k_{\sigma_1(\alpha)},l_\alpha} {U}_{k_{\sigma_2(\alpha)},l_\alpha},
\end{eqnarray}
where    matrix  $J$, the partial indistinguishability matrix, indexed by two permutations $\sigma_1$ and $\sigma_2$ of $N$ elements, reads 
\begin{eqnarray}
\label{E6}
&& J(\sigma_1,\sigma_2) = \sum_{\vec{s}} \int\rd \vec{\omega} \prod_{\alpha=1}^N \Gamma_{l_\alpha}(s_\alpha,\omega_\alpha)\nonumber\\
& & \times 
G(\{s_{\sigma^{-1}_1(\alpha)},\omega_{\sigma^{-1}_1(\alpha)}\}|\{s_{\sigma^{-1}_2(\alpha)},\omega_{\sigma^{-1}_2(\alpha)}\}).
\end{eqnarray}
 Here we note that  for different  detector sensitivities $\Gamma_{l_1},\ldots,\Gamma_{l_N}$,    matrix elements $J(\sigma_1,\sigma_2)$ also   depend on chosen output modes, thus a subscript $\vec{m}$  must be attached to them. However, for simplicity of  notations we omit it.  The matrix $J$  is Hermitian, $J^*(\sigma_1,\sigma_2)  = J(\sigma_2,\sigma_1)$, and nonnegative definite. 
 
 The $J$-matrix expansion of output probability    was first   introduced in Ref. \cite{NDBS} for $N$ single photons in mixed spectral  states   to study a model of Boson-Sampling computer with realistic sources. It  is also equivalent to   rate matrix used in more recent  Ref. \cite{SU3A}. 
Our $J$-matrix  generalizes an old observation  \cite{Mandel1991} that there is a deep relation between    indistinguishability of  photons  and   fringe visibility at output of  a beam-splitter (see details in  section \ref{subMandel}). For two photons in mixed spectral  states, a similar approach  based on  identifying a  partial indistinguishability parameter  was also used in Ref. \cite{Loudon}.

There is a  continuous family  of  spectral states of photons which correspond to the same   $J$-matrix (see also section \ref{sec2A} below) and, therefore, to  the same   probability distribution at output of a given unitary network.   Let us unite all possible output probability distributions for all possible unitary networks $U$ (for all $M$) with the same $J$-matrix    in a single class.  The question is whether a unique $J$-matrix corresponds to each such class of output probability distributions?  In other words,  are  two different setups  corresponding to two  different  $J$-matrices   distinguishable by experiments with unitary linear networks?  On the first sight, there seems to be more parameters in a $J$-matrix than one can recover from such a class of output probability distributions. Indeed,   the quadratic form of  Eq. (\ref{E5}) depends on $N!$ complex  variables, but  is evaluated at $X_\sigma \equiv \prod_{\alpha=1}^N U_{k_{\sigma(\alpha)},l_\alpha}$, i.e.,  involving at most $N^2$ independent elements of a network matrix $U$. Thus it seems that  for sufficiently large $N$ the   information  contained in  $J$-matrix  cannot be deduced   from a given  class of  output probabilities (which would require independently varying $X_\sigma$ for different   $\sigma\in \mathcal{S}_N$).  However,  note also that not every positive definite Hermitian matrix can be a $J$-matrix of a photonic input, since it must be given  according to Eq. (\ref{E6}) which imposes some  conditions,  making  the above reasoning  not conclusive.   We will not discuss this question any further in this work, relegating it to a future investigation.

Output probability of Eq. (\ref{E5}) can be also  thought of as  a multinomial,  of total power $N^2$, in  $2N^2$  matrix elements $U_{k_\alpha,l_\beta}$ and $U^*_{k_\alpha,l_\beta}$.  But this   approach, though reducing the number of used  variables,  loses the  attractive  simplicity of our approach  with  $J$-matrix with a clear physical interpretation, given above, where $X_\sigma$ serves as a ``path amplitude'' of fictitious particles (this interpretation is employed in section \ref{secZero} below for formulation of zero probability conjecture).

\textit{Auxiliary Hilbert space for spectral states.} -- To clarify the  mathematical structure of   the expressions in Eqs.~(\ref{E5})-(\ref{E6}) let us introduce an auxiliary Hilbert space $\mathcal{H}$ for description of  spectral   state of photons (a similar  method was  employed in Ref. \cite{NDBS}). Let us  denote by  $|s,\omega\rangle$    a basis   vector  for expansion of spectral state of a single particle, then
\be
\sum_{s}\int\rd \omega |s,\omega\rangle\langle s,\omega| = I.
\en{E7} 
A spectral state of $N$ particles  belongs to  the tensor product space $\mathcal{H}^{\otimes^N}$ (the auxiliary particles are distinguishable objects). A basis vector in  $\mathcal{H}^{\otimes^N}$  will be denoted by   $|\vec{s},\vec{\omega}\rangle \equiv |s_1,\omega_1\rangle\otimes\ldots\otimes|s_N,\omega_N\rangle$.
With these definitions,   a density matrix  describing  spectral state of photons   is obtained by simply replacing the Fock basis states in the expansion of   input density matrix $\rho(\vec{n})$ of Eq. (\ref{E1}) by respective tensor product states, i.e., 
\be
\hat{\rho} \equiv \sum_{\vec{s}{\,}^\pr} \sum_{\vec{s}} \int\rd \vec{\omega}^\pr\!\! \int\rd \vec{\omega} G(\vec{s}^{\,\pr},\vec{\omega}^\pr|\vec{s},\vec{\omega}) |\vec{s}{\,}^\pr,\vec{\omega}^\pr\rangle\langle\vec{s},\vec{\omega}|,
\en{E8} 
the normalization condition of Eq. (\ref{E2}) ensures that $\hat{\rho}$ has trace equal to 1 (positivity of $\hat{\rho}$  also follows from that of $\rho$ in Eq. (\ref{E1})). Permutation operations in the auxiliary space  $\mathcal{H}^{\otimes^N}$  play an essential role below.  A permutation operator $P_\sigma$, corresponding to a permutation $\sigma$ of $N$ elements,  acts in $\mathcal{H}^{\otimes^N}$ as follows 
\be
P_\sigma |j_1\rangle\otimes\ldots\otimes|j_N\rangle 
 \equiv  |j_{\sigma^{-1}(1)}\rangle\otimes\ldots\otimes|j_{\sigma^{-1}(N)}\rangle 
\en{E10}
(by this definition the vector from the $k$th Hilbert space $\mathcal{H}$ in the tensor product goes to  the $\sigma(k)$th space). The set of operators $P_\sigma$ is  a representation of the symmetric (permutation) group $\mathcal{S}_N$, i.e., we have  $P_{\sigma_1}P_{\sigma_2} = P_{\sigma_1\sigma_2}$ (note that $ P_\sigma^\d = P_{\sigma^{-1}}$). Below we will frequently refer  to  permutations $\pi$ exchanging spectral states of photons  in each    input mode between themselves, thus we associate with   the Hilbert space in position  $\alpha$  in the tensor product $\mathcal{H}^{\otimes N}$ an input mode index $k_\alpha$  of  naturally ordered set $(k_1,...,k_N)$, therefore we can identify such permutations with subgroup  $ \mathcal{S}_{n_1}\otimes...\otimes \mathcal{S}_{n_M}$ acting  on  $\mathcal{H}^{\otimes N}$. 

Due to symmetry property of $G$-function of  Eq. (\ref{E1}), we have  for any permutation $\pi \in \mathcal{S}_{n_1}\otimes...\otimes \mathcal{S}_{n_M}$ 
\be
P_\pi \hat{\rho} = \hat{\rho}P_\pi = \hat{\rho}.
\en{SymHatRho}
For instance, in case of diagonal representation, Eq. (\ref{E1A}), and  fluctuating parameter case, Eq. (\ref{E1B}) with   respective basis states $|\tilde{\Phi}(x)\rangle$ being  linearly independent (e.g.,   photons in spectral  states of a Gaussian  shape  with  fluctuating arrival times), property (\ref{SymHatRho}) implies that respective basis functions   $\Phi(x; \vec{s},\vec{\omega})$  are invariant under permutations $\pi\in \mathcal{S}_{n_1}\otimes...\otimes \mathcal{S}_{n_M}$ of $({s}_\alpha,{\omega}_\alpha)$.

Let us also introduce a detector  operator which has a diagonal representation  in the above defined auxiliary Hilbert space, i.e.,
\be
\hat{\Gamma}_l \equiv  \sum_{\vec{s}}   \int\rd \vec{\omega}\, \Gamma_l(s,\omega) |s,\omega\rangle\langle s, \omega|. 
\en{E9} 
Then  matrix $J$ defined in  Eq. (\ref{E6}) assumes  the following compact form 
\begin{eqnarray}
\label{E11}
J(\sigma_1,\sigma_2)& = & \sum_{\vec{s}}   \int\rd \vec{\omega}\, \langle \vec{s},\vec{\omega}|\hat{\Gamma}_{l_1}\otimes\ldots\otimes\hat{\Gamma}_{l_N} P^\d_{\sigma_2}\hat{\rho}P_{\sigma_1}|\vec{s},\vec{\omega}\rangle\nonumber\\
&=& \mathrm{Tr}\{\hat{\Gamma}_{l_1}\otimes\ldots\otimes\hat{\Gamma}_{l_N}P^\d_{\sigma_2}\hat{\rho}P_{\sigma_1} \},
\end{eqnarray} 
where the trace  is taken in   $\mathcal{H}^{\otimes^N}$. In its turn,  output probability  of  Eq. (\ref{E5})  becomes 
\be
P(\vec{m}|\vec{n})  =  \frac{1}{\mu(\vec{m})\mu(\vec{n})}\mathrm{Tr}\{\hat{\Gamma}_{l_1}\otimes\ldots\otimes\hat{\Gamma}_{l_N}\mathcal{U}_N \hat{\rho}\,\mathcal{U}^\dag_N \},
\en{E12}
where we have introduced an operator $\mathcal{U}_N$ acting in $\mathcal{H}^{\otimes^N}$ and given by 
\be
\mathcal{U}_N \equiv \sum_{\sigma}\left[ \prod_{\alpha=1}^N U_{k_{\sigma(\alpha)},l_\alpha} \right]P_\sigma^\d. 
\en{E13}
Though Eqs. (\ref{E11})-(\ref{E13}) are  an equivalent representation of  Eqs. (\ref{E5}) and (\ref{E6}),   the former set of equations makes it clearer how to analyze the results  by  application of the methods of linear algebra in the Hilbert space.

By definition, in case of a general (e.g., entangled) input,  $J$-matrix involves a trace in the tensor product space $\mathcal{H}^{\otimes N}$. However, one can easily show that in case of  factorized  input   (e.g., for independent photon sources),
\be
\hat{\rho} = \prod_{\alpha=1}^N{\!}^{\otimes} \hat{\rho}_{\alpha},
\en{rhoFACT}
or for an  input being a convex combination of such  factorized states the corresponding $J$-matrix is expressed through  some traces only in $\mathcal{H}$. Indeed, for an arbitrary permutation $\sigma$, by using Eq. (\ref{E10}),  we obtain the following identity between a trace in $\mathcal{H}^{\otimes N}$ and that in $\mathcal{H}$
\be
\mathrm{Tr}\left\{P^\dag_\sigma \prod_{\alpha=1}^N{\!}^{\otimes}A_\alpha \right\} = \prod_{j=1}^q\mathrm{Tr}\left\{A_{\alpha_{j1}}\ldots A_{\alpha_{j\ell_j}} \right\},
\en{IdTr}
where $c_1,\ldots,c_q$ is the set of disjoint cycles in the decomposition  $\sigma = c_1\cdot\ldots \cdot c_q$, cycle $c_i$ is assumed to be given by $\alpha_{j1}\to\alpha_{j2}\to\ldots\to\alpha_{j\ell_j}\to\alpha_{j1}$, and  $\ell_j$  is  cycle  length.  Therefore, assuming the above  cycle structure of $\sigma_R \equiv \sigma_2\sigma^{-1}_1$, for an input of Eq. (\ref{rhoFACT}) we obtain from Eq. (\ref{E11}) 
\be
J(\sigma_1,\sigma_2)  =  \prod_{j=1}^q\mathrm{Tr}\left\{ \hat{\Gamma}_{l_{\sigma^{-1}_2(\alpha_{j1}) } }\hat{\rho}_{{\alpha_{j1}} }\ldots  \hat{\Gamma}_{l_{\sigma^{-1}_2(\alpha_{j\ell_j}) } }\hat{\rho}_{{\alpha_{j\ell_j}} }\right\}.
\en{JFACT}
From Eq. (\ref{JFACT}) it is seen that  for identical detectors  and   input (\ref{rhoFACT}) $J(\sigma_1,\sigma_2)$  depends only on  the cycle decomposition of the   relative permutation $\sigma_2\sigma^{-1}_1$.

\subsection{Completely indistinguishable and maximally distinguishable photons with  ideal detectors: $J$-matrices and corresponding  inputs}
\label{sec2A}

First of all, one can easily verify  that  for  maximally efficient  detectors, $\Gamma_{l}(s,\omega) = 1$, the  output probabilities sum to 1, as it should be. Indeed,   in this case  Eqs. (\ref{E11})-(\ref{E13}) give 
\begin{eqnarray}
\label{E14}
&&\sum_{|\vec{m}|=N} P(\vec{m}|\vec{n}) = \sum_{\vec{l}} \frac{\mu(\vec{m})}{N!} \frac{1}{\mu(\vec{m})\mu(\vec{n})}\mathrm{Tr}\{\mathcal{U}_N \hat{\rho}\,\mathcal{U}^\dag_N \}\nonumber\\
&& = \frac{1}{N!\mu(\vec{n})} \sum_{\sigma_1}\sum_{\pi} \mathrm{Tr}\{P^\d_{\sigma_1}P^\d_{\pi}\hat{\rho}P_{\sigma_1} \} =  \mathrm{Tr}\{\hat{\rho}\} = 1, \qquad
\end{eqnarray}
where we have used an identity due to unitarity of  network matrix $U$
\begin{eqnarray*}
&&\prod_{\alpha=1}^N\sum_{l_\alpha=1}^M U_{k_{\sigma_2(\alpha)},l_\alpha}U^*_{k_{\sigma_1(\alpha)},l_\alpha} = \prod_{\alpha=1}^N \delta_{k_{\sigma_1(\alpha)},k_{\sigma_2(\alpha)}}\\
&& = \sum_{\pi} \delta_{\sigma_2\sigma^{-1}_1,\pi},
\end{eqnarray*}
with  $\pi \in \mathcal{S}_{n_1}\otimes...\otimes \mathcal{S}_{n_M}$  (thus   $\sum_{\pi}1 = \mu(\vec{n})$). 

Eqs. (\ref{E12}) and (\ref{E13}) generalize  the well-known fact \cite{C,S} that in the ideal case of completely indistinguishable photons and ideal  detectors  the bosonic  output probability  in an unitary linear network is expressed through the absolute value squared of the  matrix permanent of  a $N\times N$-dimensional  matrix $U[\vec{n}|\vec{m}]$, built from the  network matrix by selecting,  with repetitions, rows (columns)  corresponding to the input  $\vec{n}$ (the output $\vec{m}$)  of a considered transition, i.e.,
\be
P^{(ind)}(\vec{m}|\vec{n}) = \frac{|\sum_\sigma\prod_{\alpha=1}^NU_{k_\sigma(\alpha),l_\alpha} |^2}{\mu(\vec{m})\mu(\vec{n})} = \frac{|\mathrm{per}(U[\vec{n}|\vec{m}])|^2}{\mu(\vec{m})\mu(\vec{n})},
\en{E15}
where   the permanent of an $N\times N$-dimensional  matrix $A$ is defined as follows $\mathrm{per}(A) = \sum_\sigma \prod_{\alpha=1}^NA_{\sigma(\alpha),\alpha}$ (for a  discussion of properties of the matrix permanents, see Ref. \cite{Minc}). In this case  
\be
J^{(ind)}(\sigma_1,\sigma_2) = 1,
\en{E16}
for all  permutations $\sigma_1$ and $\sigma_2$, i.e.,    matrix  $J^{(ind)}$  (\ref{E16})  is pure (has rank 1) 
\be
J^{(ind)} = {v}^\d v, \quad v \equiv (1,\ldots,1),\quad |\vec{v}|= N!, 
\en{E17}
where $ |v|\equiv\sum_j |v_j|$. It has only one nonzero eigenvalue equal to $N!$. Now let us see what input states give $J$-matrix of Eq. (\ref{E16}). Using that  $\mathrm{Tr}\{P^\d_{\sigma_2}\hat{\rho}^{(ind)}P_{\sigma_1}\} = \mathrm{Tr}\{P_{\sigma_1\sigma^{-1}_2}\hat{\rho}^{(ind)} \} = 1$  one can establish that in  the diagonal representation following from  Eq. (\ref{E1A}), i.e., 
\be
\hat{\rho} = \sum_i p_i | {\Phi}_i\rangle\langle  {\Phi}_i|,\; | {\Phi}_i\rangle = \sum_{\vec{s}}\int\rd\vec{\omega}\,\Phi_i(\vec{s}, \vec{\omega})|\vec{s},\vec{\omega}\rangle,
\en{hatRhoDiag}
applied to $\hat{\rho}^{(ind)}$, the basis states are symmetric: $P_\sigma| {\Phi}_i\rangle =  | {\Phi}_i\rangle$ for any $\sigma\in \mathcal{S}_N$. Similar conclusion applies to  an expansion  over a basis of non-orthogonal linearly independent states, following from Eq. (\ref{E1B}).     The corresponding functions $\Phi_i(\vec{s}, \vec{\omega})$ and, hence,  $G^{(ind)}(\vec{s}{\,}^\pr,\vec{\omega}^\pr|\vec{s},\vec{\omega})$ are symmetric with respect to any permutation of  their arguments.  We note that a similar condition was  first established  in Ref. \cite{Ou2006}.  For completely indistinguishable single  photons  each basis state $| {\Phi}_i\rangle$ in the expansion of $\hat{\rho}^{(ind)}$ is of the form 
\be 
| {\Phi}_i\rangle = \frac{c_i}{N!} \sum_\sigma  \prod_{\alpha=1}^N {}^{\otimes} |\phi^{(i)}_{\sigma(\alpha)}\rangle ,
\en{1Photmode} 
where the normalization coefficient is given by $c^2_i = N!/ \mathrm{per}(\mathcal{G}^{(i)})$ with $\mathcal{G}^{(i)}_{\alpha\beta} =  \langle\phi^{(i)}_\alpha|\phi^{(i)}_\beta\rangle$.  A similar  observation  was first employed  in Ref.  \cite{Branning1999} for engineering  the   complete indistinguishability by  coherently overlapping  two   processes    for  creation of a pair of photons.

Guided by the above, we will  say that the photons  are maximally distinguishable if the respective matrix $J$  is maximally mixed as allowed by Eq. (\ref{SymHatRho}).    From Eqs. (\ref{SymHatRho}) and  (\ref{E11})  we have  for $\pi_{1,2}\in \mathcal{S}_{n_1} \otimes\ldots\otimes \mathcal{S}_{n_M}$
 \be
J(\pi_1\sigma_1,\pi_2\sigma_2)  = J(\sigma_1,\sigma_2), 
 \en{E18}
 hence, the most  mixed $J$  reads
\be
J^{(cl)}(\sigma_1,\sigma_2)  = \sum_\pi \delta_{\sigma_2,\pi\sigma_1}  = \frac{1}{\mu(\vec{n})} \sum_\pi\sum_{\pi^\pr} \delta_{\pi^\pr\sigma_2,\pi\sigma_1},
\en{E19} 
where  the second form  manifests  compliance with   the required symmetry of Eq.  (\ref{E18}).  Note  that matrix $J^{(cl)}$ has a  block-diagonal  form
\be
J ^{(cl)}= {\sum_{q}}^{\oplus} v^\d_q v_q, \quad v_q \equiv  (1,\ldots,1),\quad |{v}_q| = \mu(\vec{n}),
\en{E20}
where there are $\frac{N!}{\mu(\vec{n})}$ blocks (terms in the   direct sum).   The states in the diagonal representation (\ref{hatRhoDiag}) of $\hat{\rho}^{(cl)}$  satisfy the property 
\be
\langle  {\Phi}_i|P_{\sigma}| {\Phi}_i\rangle = 0
\en{Phicl}
  for all permutations  $\sigma \notin \mathcal{S}_{n_1}\otimes...\otimes \mathcal{S}_{n_M}$. The same property    applies to   expansion as in Eq. (\ref{hatRhoDiag})  over a basis of non-orthogonal but  linearly independent states.    In an equivalent form, this condition can be  formulated for the corresponding  function $G^{(cl)}$  as the following   orthogonality condition 
  \be
\sum_{\vec{s}} \int\rd \vec{\omega}\, G^{(cl)}(\{s_{\sigma(\alpha)},\omega_{\sigma(\alpha)}\}|
\{s_\alpha,\omega_\alpha\})=0
 \en{Gcl}
 for $ \sigma \notin \mathcal{S}_{n_1}\otimes...\otimes \mathcal{S}_{n_M}$.  A similar  condition  was first discussed  in Ref. \cite{Ou2006}.  The output probability corresponding to the $J^{(cl)}$ of Eq. (\ref{E19})  reads
\begin{eqnarray}
\label{E21}
P^{(cl)}(\vec{m}|\vec{n})& = &  \frac{\sum_{\sigma}\sum_{\pi} \prod_{\alpha=1}^NU^*_{k_{\sigma(\alpha)},l_\alpha}U_{k_{\pi\sigma(\alpha)},l_\alpha}}{\mu(\vec{m})\mu(\vec{n})} \nonumber\\
&=&   \frac{\sum_{\sigma} \prod_{\alpha=1}^N|U_{k_{\sigma(\alpha)},l_\alpha}|^2}{\mu(\vec{m})},
\end{eqnarray} 
since   for $\pi \in \mathcal{S}_{n_1}\otimes...\otimes \mathcal{S}_{n_M}$ we have   $U_{k_{\pi(\alpha)},l_\alpha} = U_{k_\alpha,l_\alpha}$. 

Let us note the following feature. The trace of  matrix $J$, i.e., $\mathrm{Tr}\{J\} =\sum_\sigma J(\sigma,\sigma)$, for ideal detectors, coincides with the  number $N!$ of different paths. For    completely indistinguishable photons,  Eq. (\ref{E17}), all paths interfere with equal weights  (see  Eq. (\ref{E15})), whereas when  photons in  different input modes are maximally distinguishable, Eq. (\ref{E20}),  there is no path interference contribution to the output  probability.  The  output probability in the latter case has a natural classical interpretation, if one assumes that  classical particles   are classically indistinguishable, i.e., if their paths through the network are not traced. In this case,  Eq. (\ref{E21}) describes  transition probability of $N$ indistinguishable  classical particles through  a Markovian network  whose  transition  matrix   element $A_{kl}$ is defined by  $A_{kl} = |U_{kl}|^2$.

\subsection{Completely indistinguishable and maximally distinguishable photons with realistic detectors }
\label{sec2B}

Let us see what changes occur  in the above two extreme cases  when realistic detectors with generally  different   efficiencies $\Gamma_l(s,\omega)$ are used. In this case  probability formula (\ref{E12}) applies to a  post selected case, when all input  photons are  detected. The trace of $J$-matrix in this case  is less than $N!$. We have 
\begin{eqnarray}
\label{E23}
&&J(\sigma,\sigma) =   \mathrm{Tr}\{\hat{\Gamma}_{l_1}\otimes\ldots\otimes\hat{\Gamma}_{l_N}P^\d_\sigma\hat{\rho}P_\sigma \}\nonumber\\
&& = \sum_{\vec{s}}\int\rd\vec{\omega}\,G(\vec{s},\vec{\omega}|\vec{s},\vec{\omega}) \prod_{\alpha=1}^N\Gamma_{l_\alpha}(s_{\sigma(\alpha)},\omega_{\sigma(\alpha)}). \qquad
\end{eqnarray}
For completely indistinguishable   photons  $J(\sigma,\sigma) $ is independent of $\sigma$ since  $G$ is completely symmetric under $\mathcal{S}_N$. 
Therefore, to reduce  this  case with  realistic detectors to  that of ideal detectors,     a single additional parameter, the detection probability $D$, 
\be
D^{(ind)} = \sum_{\vec{s}}\int\rd\vec{\omega}\,G^{(ind)}(\vec{s},\vec{\omega}|\vec{s},\vec{\omega})  
\prod_{\alpha=1}^N\Gamma_{l_\alpha}(s_\alpha,\omega_\alpha),
\en{E27}
 independent of  a considered network, must be defined.  We obtain $J$-matrix of the  form  (compare with Eq. (\ref{E17}))
 \be
J^{(ind)} = D^{(ind)}{v}^\d{v}, \quad {v} \equiv (1,\ldots,1),\quad |{v}|= N!.
\en{E25}
Output probability  is thus   multiplied by    $D^{(ind)}$.

For  maximally distinguishable photons one can use the diagonal form (\ref{hatRhoDiag})  and note that by definition in the maximally distinguishable case  $J(\sigma_1,\sigma_2)\ne0$ only for $\sigma_2\sigma_1^{-1} \in \mathcal{S}_{n_1}\otimes...\otimes \mathcal{S}_{n_M}$.  This   occurs under a condition   involving detector sensitivities (replacing Eq. (\ref{Phicl}))

\be
\langle  {\Phi}_i|\left[\prod_{\alpha=1}^N{}^{\otimes} \hat{\Gamma}_{l_{\sigma^{-1}_1(\alpha)}}\right]  P^\dag_{\sigma_2\sigma^{-1}_1}| {\Phi}_i\rangle = 0
\en{PhiclGen} 
for all permutations  satisfying $\sigma_{2}\sigma^{-1}_1 \notin \mathcal{S}_{n_1}\otimes...\otimes \mathcal{S}_{n_M}$ and $l_\alpha$ of considered transitions. Eq.~(\ref{PhiclGen}), thanks to the dependence also  on  $\sigma_1$,  places  more conditions  on spectral states of photons   than  Eq. (\ref{Phicl})  for  ideal detectors. Moreover, for  general  dissimilar  detectors, the corresponding $J(\sigma,\sigma)$  depends on $\sigma$. In matrix form  (compare with Eq.~(\ref{E20}))
\be
J^{(cl)} =  {\sum_{\tau}}^{\oplus} D^{(cl)}(\tau) v^\d_\tau v_\tau, \; v_\tau \equiv  (1,\ldots,1),\; |{v}_\tau| = \mu(\vec{n}), 
\en{E26}
where $\tau$ is permutation  of indices $\alpha$ belonging to  different input modes in  the  decomposition $ \sigma_2\sigma^{-1}_1  = \tau \pi$  with $\pi \in \mathcal{S}_{n_1}\otimes...\otimes \mathcal{S}_{n_M}$ and 
\be
D^{(cl)}(\tau) = \sum_{\vec{s}}\int\rd\vec{\omega}\,G^{(cl)}(\vec{s},\vec{\omega}|\vec{s},\vec{\omega}) \prod_{\alpha=1}^N\Gamma_{l_\alpha}(s_{\tau(\alpha)},\omega_{\tau(\alpha)}).
\en{Dcl}

 The above two examples imply that one has to be careful in attributing a nearly zero output probability to  quantum interference (for  nonzero probability of  single particle transition), since it may well happen that  the zero probability is due to some generalization of the above defined  detection factors $J(\sigma,\sigma)\ll1$, present in the  maximally distinguishable (classical) case as well. A specific case of Gaussian-shaped single photons with different arrival times is considered in appendix \ref{appC}.  For arbitrary detectors and arbitrary input   Eq. (\ref{E1}) we  introduce a reduced   $J$-matrix   in section \ref{sec2C} below.

 \subsection{Output probability in  terms of  the matrix  permanents} 
 \label{secPER}
 
 Let us establish the form  of output probability in the  general case of arbitrary input of Eq. (\ref{E1}). We employ the diagonal representation (\ref{hatRhoDiag}). Output probability Eq. (\ref{E12}) can be also cast as 
 \be
 P(\vec{m}|\vec{n}) = \frac{1}{\mu(\vec{m})\mu(\vec{n})} \sum_i p_i \langle \Psi^{(i)}| \Psi^{(i)} \rangle,
 \en{EQ1}
 where we have introduced $| \Psi^{(i)}\rangle \in \mathcal{H}^{\otimes N}$ as follows
 \be
 |\Psi^{(i)}\rangle \equiv \sum_\sigma \left[\prod_{\alpha=1}^N\!{}^{\otimes} U_{k_{\sigma(\alpha)},l_\alpha}\sqrt{\hat{\Gamma}_{l_\alpha}}\right]
 P^\dag_{\sigma}|\Phi_i\rangle. 
 \en{EQ2}
 Let us use an  orthogonal basis $|j\rangle$  in the Hilbert space $ \mathcal{H}$ and expand 
  \be
 |\Phi_i\rangle = \sum_{\vec{j}} C^{(i)}_{\vec{j}} |\vec{j}\rangle, 
 \en{EQ3}
 where   $|\vec{j}\rangle = |j_1\rangle\otimes\ldots\otimes |j_N\rangle \in \mathcal{H}^{\otimes N}$. 
From Eqs. (\ref{EQ2})  and (\ref{EQ3}) we obtain 
 \begin{eqnarray}
 \label{EQ4}
&& \langle\vec{j}|\Psi^{(i)}\rangle = \sum_{\vec{j}^\pr}C^{(i)}_{\vec{j}^\pr}\sum_\sigma \prod_{\alpha=1}^N U_{k_{\sigma(\alpha)},l_\alpha}\langle j_\alpha|\sqrt{\hat{\Gamma}_{l_\alpha}}|j^\pr_{\sigma(\alpha)}\rangle\nonumber\\
&& =  \sum_{\vec{j}^\pr}C^{(i)}_{\vec{j}^\pr} \mathrm{per}\left(U[\vec{n}|\vec{m}]\cdot B(\vec{j},\vec{j}^\pr)\right),
 \end{eqnarray}
 here (and throughout the text)   the central dot in a product of two matrices denotes the Hadamard (entry-wise) product, in this case of   matrix $U[\vec{n}|\vec{m}]$   (built, as above described, from  network matrix $U$) and   matrix $B(\vec{j},\vec{j}^\pr)$ defined  as follows
 \be
 B_{\beta,\alpha} (\vec{j},\vec{j}^\pr) \equiv \langle j_\alpha|\sqrt{\hat{\Gamma}_{l_\alpha}}|j^\pr_{\beta}\rangle. 
 \en{EQ5}
 Using Eq. (\ref{EQ4}) into Eq. (\ref{EQ1}) we obtain the result 
 \be
 P(\vec{m}|\vec{n}) = \frac{1}{\mu(\vec{m})\mu(\vec{n})} \sum_i p_i \sum_{\vec{j}} \biggl| \sum_{\vec{j}^\pr}C^{(i)}_{\vec{j}^\pr} \mathrm{per}\left(V(\vec{j},\vec{j}^\pr)\right) \biggr|^2
 \en{EQ6}
 with $V(\vec{j},\vec{j}^\pr) \equiv U[\vec{n}|\vec{m}]\cdot B(\vec{j},\vec{j}^\pr)$. 
 
One can use any  basis of tensor product states  for expansion in Eq.  (\ref{EQ3}), for instance, in the standard spectral basis $|\vec{s},\vec{\omega}\rangle$ we have 
\begin{eqnarray}
\label{EQ7}
 && P(\vec{m}|\vec{n})= \frac{1}{\mu(\vec{m})\mu(\vec{n})} \sum_i p_i \sum_{\vec{s}}\int\rd\vec{\omega}\nonumber\\
 & & \times \biggl| \sum_{\vec{s}{\,}^\pr}\int\rd \vec{\omega}^\pr \Phi_i(\vec{s}{\,}^\pr,\vec{\omega}^\pr)\mathrm{per}\Bigl(V(\vec{s},\vec{\omega},\vec{s}{\,}^\pr,\vec{\omega}^\pr)\Bigr) \biggr|^2,\qquad
\end{eqnarray}
where $V(\vec{s},\vec{\omega},\vec{s}{\,}^\pr,\vec{\omega}^\pr) = U[\vec{n}|\vec{m}] \cdot B(\vec{s},\vec{\omega},\vec{s}{\,}^\pr,\vec{\omega}^\pr)$ with 
\be
B_{\beta,\alpha} (\vec{s},\vec{\omega},\vec{s}{\,}^\pr,\vec{\omega}^\pr)\equiv \delta_{s^\pr_\beta,s_\alpha}\delta(\omega^\pr_\beta-\omega_\alpha)
\left[\Gamma_{l_\alpha}(s_\alpha,\omega_\alpha)\right]^{\frac12}.
\en{EQ8}
 For example,  Eq.~(\ref{EQ7}) simplifies to Eq.~(\ref{E21}) for ideal detectors if, using  the definition of $B$-matrix (\ref{EQ8}), one first   integrates (sums)  over $\vec{\omega}$ ($\vec{s}$) in Eq.~(\ref{EQ7})  by using orthogonality condition    (\ref{Phicl}),  i.e., $\sum_{\vec{s}}\int\rd\vec{\omega} {\Phi_i}^*(\vec{s}{\,},\vec{\omega})\Phi_i(\{s_{\sigma(\alpha)},\omega_{\sigma(\alpha)}\}) = \delta_{\sigma,\pi}$ where \mbox{$ \pi \in \mathcal{S}_{n_1} \otimes\ldots\otimes \mathcal{S}_{n_M}$}. The  result is nothing but the  $J$-matrix representation (\ref{E5}) with $J$ of Eq. (\ref{E19}) which can be  evaluated further   according to calculation of section \ref{sec2A}, see  Eqs. (\ref{E19}) and (\ref{E21}). 
 
 One final observation is in order. In Eqs. (\ref{EQ6}) or (\ref{EQ7}) the squared absolute value is taken of a coherent sum of the matrix  permanents.   In case of single photons from  independent  sources, i.e., when the input density matrix  is given by Eq. (\ref{rhoFACT}) with each $\hat{\rho}_\alpha$ being a density matrix in $\mathcal{H}$, one can also express output probability as sum (or integral) over absolute values squared of the matrix permanents   by using a different  matrix for  spectral data in the Hadamard product (see sections \ref{sec3A} and \ref{sec3B} below).

  \subsection{$J$-matrix based measure of quantum coherence of photons}
\label{sec2C}

We have found above the form of   $J$-matrix in the extreme cases of completely indistinguishable and maximally distinguishable photons for arbitrary detectors. Taking into account these results, it is suggestive to look for  a $J$-matrix based measure of  quantum coherence of a multi-photon input for a given   set of detectors.  Note that   quantum coherence of  photon paths  is reflected in $J$-matrix  in a way very similar as it would be in a  usual density matrix of a quantum system  (with the exception of  the  normalization). Using this observation, below we propose to use the purity as a measure of coherence  for photons, which generalizes Mandel's parameter \cite{Mandel1991} for $N>2$. This measure is also a measure of partial indistinguishability, similar as it is in Mandel's case of two photons. We consider an arbitrary $M$-mode network  given by some unitary matrix $U$. 

\subsubsection{Mandel's degree of indistinguishability for  two photons}
\label{subMandel}
To begin with,  let us first consider  two-photon case studied in Ref. \cite{Mandel1991} (and after that  also in Ref. \cite{Loudon}) where it was found that a single parameter is both a degree of indistinguishability and   a degree of quantum coherence   (how the   degree of indistinguishability    depends  on  different  parameters in   spectral  states of photons is recently studied in Ref. \cite{Falk2012}).  

For  two single photons in spectral states $\hat{\rho}_1$ and $\hat{\rho}_2$ at   input modes $k_1\ne k_2$ we have  only two permutations $\sigma = I$ (trivial) and  $\sigma=T$ (transposition of two photons). From Eq. (\ref{E11}), by  using the properties  $P_t\hat{\Gamma}_{l_1}\otimes\hat{\Gamma}_{l_2}P_t = \hat{\Gamma}_{l_2}\otimes\hat{\Gamma}_{l_1}$ and $\mathrm{Tr}(A\otimes BP_t) = \mathrm{Tr}(AB)$ (where the latter trace is in $\mathcal{H}$, whereas the former is in $\mathcal{H}\otimes \mathcal{H}$), we obtain 
\begin{eqnarray}
\label{JN2}
&& J(I,I) = \mathrm{Tr}(\hat{\Gamma}_{l_1}\hat{\rho}_1) \mathrm{Tr}(\hat{\Gamma}_{l_2}\hat{\rho}_2), \nonumber\\
&& J(T,T) = \mathrm{Tr}(\hat{\Gamma}_{l_2}\hat{\rho}_1) \mathrm{Tr}(\hat{\Gamma}_{l_1}\hat{\rho}_2), \nonumber\\
&& J(T,I) = \mathrm{Tr}(\hat{\Gamma}_{l_1}\hat{\rho}_1\hat{\Gamma}_{l_2}\hat{\rho}_2) = J^*(I,T).
\end{eqnarray}
Detectors    reduce the total probability of detection. Let us first try  to factor this  effect of detectors from  their   influence   on  quantum  coherence of photons.   By introducing a diagonal matrix $D(\sigma_1,\sigma_2) = \delta_{\sigma_1,\sigma_2} J(\sigma_1,\sigma_1)$ let us define   a reduced $\hat{J}$-matrix as follows 
\be
\hat{J} \equiv D^{-\frac12} J D^{-\frac12}  =  \left( \begin{array}{cc} 1& \mathcal{V}^* \\ \mathcal{V} &1\end{array}\right),
\en{Jnew}
where 
\be
\mathcal{V} \equiv \frac{J(T,I)}{\sqrt{J(I,I)J(T,T)}}.
\en{Vis}
Now, it is easy to see that $\mathcal{V}$ is exactly Mandel's indistinguishability  parameter \cite{Mandel1991}, whose absolute value  gives the strength of coherence for two  photons. Indeed, if both photons are detected, then  the defined $\hat{J}$-matrix  describes  their indistinguishability. It has the correct trace and,  since the original matrix $J$ is Hermitian and positive definite, $|\mathcal{V}|\le 1$. Following \cite{Mandel1991} we expand (setting $\mathcal{V} = |\mathcal{V}|e^{i\theta}$)
\be
\hat{J} =P_{ID}{J}_{ID} + P_{D}\mathrm{diag}(1,1), \;  J_{ID} = \left(\begin{array}{cc}1 & e^{-i\theta}\\  e^{i\theta}& 1 \end{array}\right),
\en{expJ}
where $J_{ID}$ is a $J$-matrix corresponding to completely indistinguishable photons   and arbitrary detectors (if detectors are identical  $\mathcal{V}$ is real) 
with probability  $P_{ID} = |\mathcal{V}|$   and the identity matrix  corresponds to maximally distinguishable photons.  Moreover, from Eqs. (\ref{JN2}) and (\ref{Vis})  we    obviously  get $\mathcal{V} = 1$ for  $\hat{\rho}_1=\hat{\rho}_2 = |\phi\rangle\langle\phi|$, for arbitrary $|\phi\rangle$.

\subsubsection{Degree of indistinguishability for $N\ge2$}
\label{subPurity}

Guided by the examples of sections \ref{sec2A}, \ref{sec2B}, and \ref{subMandel}, we propose to use  a normalized purity $0\le \mathcal{P}\le 1$   of the  reduced $\hat{J}$-matrix as a measure of partial indistinguishability of photons. We define the normalized purity as 
\be
\mathcal{P} \equiv \frac{N!}{N!-1}\left(\mathrm{Tr}\left\{\left(\frac{\hat{J}}{N!}\right)^2\right\} - \frac{1}{N!}\right),
\en{Purity}
since $\mathrm{Tr}\{\hat{J}\} = N!$ and   matrix $\hat{J}$ is  $(N!\times N!)$-dimensional. In Mandel's case Eq. (\ref{Jnew}) we obtain $\mathcal{P} = |\mathcal{V} |^2$. 

 Similarly as in the two-photon case, for  $N$ photons we define  a $\hat{J}$-matrix   by rescaling the $J$-matrix  by its diagonal part  
 \be
\hat{J}(\sigma_1,\sigma_2) = \frac{J(\sigma_1,\sigma_2)}{\sqrt{J(\sigma_1,\sigma_1)}\sqrt{J(\sigma_2,\sigma_2)}}.
\en{hatJ}
 The necessary property $|\hat{J}(\sigma_1,\sigma_2)| \le 1$  follows from  positivity of $J$-matrix by using the Sylvester criterion.  Output probability becomes 
 \be
 P(\vec{m}|\vec{n}) = \frac{1}{\mu(\vec{m})\mu(\vec{n})} X^\dag  \hat{J}  X,
 \en{P_Z}
 where  a column-vector $X$  has elements, indexed by  $\sigma\in \mathcal{S}_N$, equal to the path   amplitudes  reduced by  detectors
 \be
 X_\sigma = \sqrt{J(\sigma,\sigma)}\prod_{\alpha=1}^N U_{k_{\sigma(\alpha)},l_\alpha}.
 \en{Z}
This transformation can be easily understood by referring  to the classical case, where $|X_\sigma|^2$ is  probability of a transition of \textit{distinguishable} classical particles in a Markovian network with losses of particles due to imperfect detections. 
 
Though, in general,  there  is  no density matrix resulting in $\hat{J}$-matrix (\ref{hatJ})   by Eq. (\ref{E11}) with ideal detectors,  it is possible to sometimes   consider   the  effect  of general detectors in a way mathematically  equivalent  to the case of ideal detectors   by adopting   a generalized inner product in the auxiliary Hilbert space  $\mathcal{H}^{\otimes N}$  in the trace-definition of $J$-matrix  Eq. (\ref{E11}) with the detector-dependent kernel 
\be
\hat{K}_{\vec{l}}\equiv \prod_{\alpha=1}^N{\!}^{\otimes}\hat{\Gamma}_{l_\alpha}, 
\en{K}
specific to a  considered output configuration.  For instance, this  approach is employed in discussion of the zero probability conjecture     in    section \ref{secZero} below. 

In section \ref{sec3C} we analytically compute   purity (\ref{Purity})  for a  model of realistic  Boson-Sampling computer  with partially distinguishable single photons.

%%%%%%%%%%%%%%%%%%%%%%%%%%%%%%%%%%%%%%%%%%%%%%%%%
\section{ Input consisting of  one   photon or vacuum per input mode}
\label{sec3}

The case of   input consisting of a photon or vacuum per   input mode can be analyzed in considerable detail in the most general form, i.e., for  arbitrary detector efficiencies and photonic spectral states. Moreover, in this case  a considerable simplification of the resulting formulae is possible, which elucidates the effect of partial indistinguishability of photons on  output probabilities. This case is also of much importance in view of the recent  proposal of the Boson-Sampling computer \cite{AA}. 

\subsection{Single photons in pure spectral states}
\label{sec3A}

Consider an  input  (\ref{E8}) corresponding to single photons in pure spectral states.  In this case the  density matrix factorizes   
\be
\hat{\rho} = \hat{\rho}_{1}\otimes\ldots\otimes\hat{\rho}_{N}, \quad \hat{\rho}_{\alpha} = |\phi_{\alpha}\rangle\langle\phi_{\alpha}|,
\en{E28}
where
\be
|\phi_\alpha\rangle = \sum_s\int\rd\omega\, \phi_\alpha(s,\omega)|s,\omega\rangle.
\en{E29}
The  partial indistinguishability matrix $J$ (\ref{E11})  becomes 
\be
 J(\sigma_1,\sigma_2) = \prod_{\alpha=1}^N \langle\phi_{{\sigma_1(\alpha)}}|\hat{\Gamma}_{l_\alpha}|\phi_{{\sigma_2(\alpha)}}\rangle
\en{E30}
where we have used Eq. (\ref{E10}). One   feature of Eq. (\ref{E30}) should be noted: Since  detector operator  $\hat{\Gamma}_l$ enters between two spectral states  in  Eq. (\ref{E30}), one can simply project it   on the subspace spanned by the spectral states of photons, i.e., use instead the operator   $\hat{\Gamma}^\prime_l  \equiv Pr\hat{\Gamma}_lPr$, where a minimum rank projector $Pr$  is such that $Pr|\phi_\alpha\rangle = |\phi_\alpha\rangle$ for each  spectral  state $|\phi_\alpha\rangle$ at network input.  Below this is implicitly assumed. This observation  simply restates  our physical  intuition   that detectors do not increase the dimension of  the linear subspace required to describe   spectral states of photons.

For identical detectors $\hat{\Gamma}_l = \hat{\Gamma}$,  from section \ref{sec2} (see Eq. (\ref{JFACT})) we know that a $J$-matrix  corresponding to input  of Eq. (\ref{E28})  actually depends only on the   cycle decomposition of the relative permutation $\sigma_R\equiv\sigma_2\sigma^{-1}_1$. We get  
 \be 
J(\sigma_1,\sigma_2)  = \prod_{j=1}^q \prod_{i=1}^{\ell_j}\langle \phi_{{\alpha_{j,i}}}|\hat{\Gamma}|\phi_{{\alpha_{j,i+1}}}\rangle,
 \en{E30A} 
where the relative permutation is decomposed into  disjoint cycles, $\sigma_R = c_1\cdot\ldots\cdot c_q$, and it is assumed that  cycle $c_j$ is 
$\alpha_{j,1}\to \alpha_{j,2} \to \ldots \to \alpha_{j,\ell_j}\to \alpha_{j,1}$ (i.e., $\ell_j+1 \equiv 1$).

 Let us   give a reduced  form  of  output probability. From Eqs. (\ref{EQ2}) and (\ref{EQ6}) we obtain  (we omit the  input argument $\vec{n}$,  $n_k\le 1$, for simplicity)
\be
P(\vec{m}) = \frac{\langle \Psi|\Psi\rangle}{\mu(\vec{m})},  
\en{E31}
where   
\be
|\Psi\rangle \equiv \sum_\sigma \prod_{\alpha=1}^N\!{}^{\otimes} U_{k_{\sigma(\alpha)},l_\alpha}\sqrt{\hat{\Gamma}_{l_\alpha}}|\phi_{{\sigma(\alpha)}}\rangle. 
\en{E32}
Components of   $|\Psi\rangle$   in the  basis $|\vec{s},\vec{\omega}\rangle$ are given as the  matrix permanents of  an $N\times N$-dimensional  matrix $V(\vec{s},\vec{\omega})$ with  elements 
\be
V_{\beta,\alpha}(\vec{s},\vec{\omega}) =  U_{k_\beta,l_\alpha}\phi_{\beta}(s_\alpha,\omega_\alpha)\left[\Gamma_{l_\alpha}(s_\alpha,\omega_\alpha)\right]^\frac12. 
\en{E33}
Indeed, we have
\begin{eqnarray}
&& \langle\vec{s},\vec{\omega}|\Psi\rangle = \sum_\sigma\prod_{\alpha=1}^NU_{k_{\sigma(\alpha)},l_\alpha}\left[{\Gamma}_{l_\alpha}(s_\alpha,\omega_\alpha)\right]^{\frac12}\langle{s_\alpha,\omega_{\alpha}}|\phi_{{\sigma(\alpha)}}\rangle\nonumber\\
&& = \mathrm{per}\bigl( V(\vec{s},\vec{\omega})\bigr).
\label{perV}
\end{eqnarray}
Matrix   $V$ is a Hadamard product  \mbox{$V(\vec{s},\vec{\omega})  = U[\vec{n}|\vec{m}]\cdot S(\vec{s},\vec{\omega})$} (instead of using  the above $B$-matrix  (\ref{EQ8}), in  case of input with at most one  photon per mode we can incorporate  spectral states of photons into  a new  matrix $S$),
where matrix $S$ reads 
\be
S_{\beta,\alpha}(\vec{s},\vec{\omega})  \equiv \phi_{\beta}(s_\alpha,\omega_\alpha)\left[\Gamma_{l_\alpha}(s_\alpha,\omega_\alpha)\right]^\frac12
\en{S}
(column $\alpha$ of $S$ depends on the spectral data $(s_\alpha,\omega_\alpha)$,  where each   entry   is equal to  spectral state of a photon  multiplied by  the square root of  spectral  sensitivity of a detector taken at $(s_\alpha,\omega_\alpha)$).  In terms of the matrix function $V(\vec{s},\vec{\omega}) $ Eq. (\ref{E31})  becomes
\be
P(\vec{m}) = \frac{1}{\mu(\vec{m})}\sum_{\vec{s}}\int\rd\vec{\omega}\,|\mathrm{per}(V(\vec{s},\vec{\omega}))|^2.
\en{E35}
 
Instead of using the natural spectral  basis $(s,\omega)$ for  expansion of  spectral state of a photon, one can employ  any other basis, which is judged more  suitable for some reason. Indeed, given   $N$ spectral states of   photons   (for arbitrary  detectors)  one needs at most $N$ basis states (however, different basis states for different setups).  Let $|1\rangle,\ldots,|r\rangle$,  with $r\le N$,  be the required  basis set. Denoting   $|\vec{j}\rangle = |j_1\rangle\otimes\ldots\otimes|j_N\rangle$ we get 
\begin{eqnarray}
 \langle \vec{j}|\Psi_{\vec{m}}\rangle &=& \sum_\sigma \prod_{\alpha=1}^N U_{k_{\sigma(\alpha)},l_\alpha} \langle j_\alpha|\sqrt{\hat{\Gamma}_{l_\alpha}}|\phi_{{\sigma(\alpha)}}\rangle
\nonumber\\
&=& \mathrm{per}\left(U[\vec{n}|\vec{m}]\cdot S(\vec{j})\right) \equiv\mathrm{per}\bigl(V(\vec{j})\bigr),
\label{Psi_m}
\end{eqnarray}
where  $V(\vec{j}\,)= U[\vec{n}|\vec{m}]\cdot S(\vec{j}\,)$ with the following  matrix $S (\vec{j}\,)$ 
\be
S_{\beta,\alpha}(\vec{j}\,) \equiv \langle j_\alpha|\sqrt{\hat{\Gamma}_{l_\alpha}}|\phi_{\beta}\rangle.
\en{Sj}
In this case, the integral of Eq. (\ref{E35}) becomes a finite sum of at most $\frac{(N+r-1)!}{N!(r-1)!}$  terms (recall that $r$ is the rank  of  a given  set of  spectral states of $N$ bosons) 
\be
P(\vec{m}) = \frac{1}{\mu(\vec{m})} \sum_{\vec{j}} \Bigl|\mathrm{per}\bigl(V(\vec{j}\,)\bigr)\Bigr|^2. 
\en{P_new}
One observation is in order. The matrix form to represent  spectral data $S_{\beta,\alpha}$, $\alpha,\beta=1,\ldots,N$ is visually attractive, however  one should keep in mind that probability is given by a matrix permanent which does not change under permutation of rows or  columns of $U_{k_\beta,l_\alpha}$ and  $S_{\beta,\alpha}$, i.e,  when such a  permutation is   applied simultaneously to both matrices. For instance, permutation $\sigma$, applied to input states  of $S$ and $U$ (row indices),  can be transferred to basis states of $S$ and output states in $U$ (column indices). This is used below for physical interpretation of the results.

From Eqs. (\ref{E35}) or (\ref{P_new})  it follows that  the zero output probability condition    of Ref. \cite{SU3,SU3A},  given as a linear combination of the  matrix permanent, the determinant, and a generalization to nontrivial group characters, called the matrix immanants, can be replaced by a   condition  involving only permanents: $\mathrm{per}(V(\vec{s},\vec{\omega}))=0$ or $\mathrm{per}\bigl(V(\vec{j}\,)\bigr)=0$ (in the latter case the  basis is arbitrary).   
  
When  photons  are completely  indistinguishable,  detectors being  identical, the matrix $S$  of Eq. (\ref{S}) has all elements equal to  some  function $f(\vec{s},\vec{\omega})$ and $S(\vec{j}\,)$ of Eq. (\ref{Sj})  has all its  elements equal to 1 (we have $r=1$  and set $|1\rangle = |\phi\rangle$). In this case  Eqs. (\ref{E35}) or  (\ref{P_new})  reduce to a single   matrix permanent of   $U_{k_\alpha,l_\beta}$. Single photons with slightly different  spectral states or slightly dissimilar detectors destroy this trivial  factorization.  However, it turns out that  zero output probability can occur in some cases  when input contains,  besides a subset of completely indistinguishable,  also only  partially indistinguishable photons. One possibility is  when  $N-1$ photons are completely indistinguishable in some spectral state $|\varphi_1\rangle$  and the $N$th photon is  in any other spectral state $|\varphi_2\rangle$. The output probability is zero for some   configurations of input and output when the network matrix is a Fourier matrix~\cite{Tichy2014}. Understanding such  cases is  important for  generalization of the HOM effect \cite{HOM} to multi-photon interference, which   could serve also for a  conditional verification of the Boson-Sampling computer \cite{VerBS}. We will study such cases  in detail in section \ref{secZero}, where we formulate a conjecture about  zero output probability.

%%%%%%%%%%%%%%%%%%%

\subsection{Single photons in mixed spectral states}
\label{sec3B}

We have considered single photons  with pure spectral states, however, this is an  unrealistic idealization.  Let us therefore   generalize the above results to single photons in arbitrary  mixed spectral states. In this case the input state $\hat{\rho}$ of Eq. (\ref{rhoFACT}) consists of  
 \be
 \quad \hat{\rho}_{\alpha} = \int\rd x\, p_\alpha(x)|\phi_\alpha(x)\rangle\langle\phi_\alpha(x)|,    \quad p_\alpha(x)\ge 0, 
\en{E47}
where   $\langle s,\omega|\phi_\alpha(x)\rangle = \phi_\alpha(s,\omega;x)$ and  $\int\rd x\, p_\alpha(x)=1$. 
One can  interpret   the state (\ref{E47}) as given by a source with parameter $x$ fluctuating according to the probability $p_k(x)$ (in general, no orthogonality condition on vectors is imposed). Several fluctuating parameters  is a trivial extension of Eq. (\ref{E47}). The corresponding partial indistinguishability matrix $J$ is a generalization of that in 
Eq.~(\ref{E30})   
\begin{eqnarray}
\label{E48}
&& J(\sigma_1,\sigma_2) = \int\rd x_1p_1(x_1)\cdot\ldots\cdot\int\rd x_N p_N(x_N)  \nonumber\\
& & \quad \times \prod_{\alpha=1}^N   \langle\phi_{{\sigma_1(\alpha)}}(x_{\sigma_1(\alpha)})|\hat{\Gamma}_{l_\alpha}|\phi_{{\sigma_2(\alpha)}}(x_{\sigma_2(\alpha)})\rangle. 
\end{eqnarray}
Therefore, the corresponding  output probability is an obvious generalization of that in Eq. (\ref{E31}) 
\begin{eqnarray}
\label{E49}
P(\vec{m})& =&\frac{1}{\mu(\vec{m})} \int\rd x_1p_1(x_1)\cdot\ldots\cdot\int\rd x_N p_N(x_N) \nonumber\\
&\times &   \langle\Psi(\vec{x})|\Psi(\vec{x})\rangle,
\end{eqnarray} 
with $\vec{x}\equiv (x_1,\ldots,x_N)$ and 
\be
|\Psi(\vec{x})\rangle \equiv \sum_\sigma \prod_{\alpha=1}^N\!{}^{\otimes} U_{k_{\sigma(\alpha)},l_\alpha}\sqrt{\hat{\Gamma}_{l_\alpha}}|\phi_{{\sigma(\alpha)}}(x_{\sigma(\alpha)})\rangle. 
\en{E50}
In this case the corresponding matrix  $V$, the Hadamard product of spectral data and network matrix,  also depends on  fluctuating parameters $x_1,\ldots,x_N$ and the expression for output  probability similar to that of Eq. (\ref{E35}) or Eq. (\ref{P_new}), depending on the chosen  basis, involves also an averaging over these fluctuating  parameters.  We note here that   the above formulae   can be generalized in a similar way to account for  detectors with   fluctuating spectral sensitivities.

%%%%%%%%%%%%%%%%%%%%%%%%%%%%%%
\subsection{Zero output probability}
\label{secZero}

Now let us analyze   zero output probability  which occurs  in some cases  of only  partially indistinguishable photons, when the network matrix is a Fourier matrix~\cite{Tichy2014}. 
The physical  meaning  of a zero output probability  with  only  partially indistinguishable photons can be established by   answering  the following question: Is  there   an \textit{exact} cancellation  of   path amplitudes of not completely indistinguishable photons?   In view of the  connection with  duality of  the which-way information and the  interference visibility, noted  in section \ref{sec2},   one would rule out such a possibility (recall that the exact HOM dip \cite{HOM}  with  two photons is used for asserting  their complete indistinguishability). Let us  consider few examples below.

\subsubsection{$N$-photons with each photon pair  in linearly independent or coinciding spectral states}
\label{sub1}

With the aim to   answer the  above question, let us analyze the examples of Ref. \cite{Tichy2014}  in more detail using our approach (we consider  photons in pure spectral states and  ideal detectors,  $\Gamma_l =1$, for a while).  Let us first consider \mbox{$N-1$} photons in a spectral state $|\varphi_1\rangle$ and an photon in a different spectral state 
$|\varphi_2\rangle$ (not necessarily orthogonal to $|\varphi_1\rangle$). It is convenient to employ the  dual basis  of non-orthogonal states  $\langle 1|,\langle 2|$, i.e., $\langle j|\varphi_i\rangle = \delta_{ij}$. One can easily verify that in  the linear  span of  spectral states of photons, subspace of $\mathcal{H}$, 
\be
\sum_{j,l=1,2} |j\rangle\langle\varphi_j|\varphi_l\rangle\langle l| = I,
\en{dual} 
thus  an expansion similar to that of Eq. (\ref{P_new}) will contain non-diagonal quadratic form with the Gram matrix $\langle\varphi_j|\varphi_l\rangle$. 

We first employ the approach based on   $S$-matrix (\ref{Sj}) and then show that the same result rather  naturally follows from  the form of   $J$-matrix~(\ref{E30}). Setting   row order  for $S$-matrix of Eq.~(\ref{Sj}) by arranging  the basis vectors as  $(\langle 1|,\ldots,\langle 1|, \langle 2|)$  we get   that $S(\vec{j})$-matrices which result in a nonzero contribution to probability in Eq. (\ref{P_new}) correspond to $\vec{j}$ being a permutation of $(1,\ldots,1,2)$. Such an $S$-matrix  reads
\be
S(\vec{j}) = \mathcal{M}(\vec{j})\left(\begin{array}{cccc} 
1& \ldots& 1 & 0\\
\vdots & & \vdots & \vdots\\
1& \ldots& 1 & 0\\ 0& \ldots&0 & 1
\end{array}\right)= \mathcal{M}(\vec{j})(v^\dag v\oplus 1), 
\en{Snew}
where $v = (1,\ldots,1)$, $|v| = N-1$, whereas $\mathcal{M}(\vec{j})$ is the matrix representation of a permutation  $\tau$ induced by a choice of basis vector  $\langle\vec{j}| = [\langle 1|\otimes\ldots\otimes\langle 1|\otimes\langle2|]P_\tau$,   i.e., $\mathcal{M}_{kl} = \delta_{l,\tau(k)}$.  Note that permutations between indistinguishable photons do not induce any change in the  matrix $S$, thus distinct matrices $S_\alpha$ correspond to  $N-1$   transpositions $\tau_\alpha=(\alpha,N)$, $\alpha=1,\ldots,N-1$, between  each pair of   photons in states $|\varphi_1\rangle$ and $|\varphi_2\rangle$ and one for the  identity permutation.  Due to the block-matrix structure of $(v^\dag v\oplus 1)$,  for each such  matrix $S_\alpha$  the matrix permanent $\mathrm{per}\left(U[\vec{n}|\vec{m}]\cdot S_\alpha\right)$ factorizes into a product of two amplitudes, one corresponding to the $N-1$ indistinguishable photons  and an amplitude corresponding to the  $N$th  photon. To get a  clear physical  interpretation of the result,  we will  transfer permutations  to  column indices, i.e., to $l_\alpha$ in $U$ and to  $j_\alpha$ in $S$.  Due to nonorthogonality of the dual basis the output probability is given as a quadratic form of such matrix permanents. With these observations, setting also  $|\vec{\varphi}\rangle = [|\varphi_1\rangle]^{\otimes(N-1)}\otimes|\varphi_2\rangle$, we obtain   (see Eq. (\ref{dual})) 
\be 
P(\vec{m}) = \frac{1}{\mu(\vec{m})}\sum_{\alpha,\beta=1}^N \langle\vec{\varphi}|P^\dag_{(\alpha,N)} P_{(\beta,N)}|\vec{\varphi}\rangle Y^*_\alpha Y_\beta
\en{P_Snew}
where $Y_\alpha = U_{k_N,l_{\alpha}} \mathrm{per}(U[\vec{n}-\vec{1}_N|\vec{m}-\vec{1}_{l_\alpha})$. Here we have defined a vector $\vec{1}_{j}$ with  only  one nonzero entry, equal to 1, in a row (column)  with  index $j$ (we  subtract  one particle in   a mode  with index $j$ from the corresponding input (output) configuration).  Now, due to linear independence of  vectors $P_{(\alpha,N)}|\vec{\varphi}\rangle$, for $\alpha=1,\ldots,N$,  a zero output probability in Eq. (\ref{P_Snew})  occurs only if   $Y_\alpha=0$ for all $\alpha=1,\ldots,N$. This condition (besides a trivial case of some $U_{kl}=0$) obviously does not involve interference of paths of distinguishable photons (i.e., it does not depend on such interference). 
 We summarize: A zero output probability in the  case described by (\ref{P_Snew})   is formulated as  \textit{an exact  cancellation of  paths of only  completely indistinguishable photons}.   We note here that, in the considered example, zero probability  requires that $N$ different quantum amplitudes of indistinguishable photons are equal to zero, which occurs with the Fourier matrices and special input modes  \cite{MPI,Tichy2014}.

 One can generalize the above result (on which path interference is responsible for an exact   zero probability)  to $Q\ge 2$   groups of photons, where group $q$  consists of photons in spectral state $|\varphi_q\rangle$,  the spectral states $|\varphi_1\rangle,\ldots,|\varphi_Q\rangle$ being  linearly independent. In this case the corresponding matrix $S(\vec{j})$, resulting in a nonzero output probability (see more details in appendix  \ref{appB}),  is a product of a permutation matrix $\mathcal{M}(\vec{j})$ and a matrix equal to a direct sum of matrices with each entry being equal to $1$:
 \be
 S(\vec{j}) = \mathcal{M}(\vec{j})\left(\sum_{q=1}^Q {\!}^{\oplus} v^\dag_q v_q\right), \; v_q \equiv (1,\ldots,1),\; |v_q| = c_q,
 \en{S_gen}
 where $c_q$ is the number of photons in spectral state $|\varphi_q\rangle$.     A notable feature of this case is that path interference of   photons  within   each group   is maximally possible. Note that  photons in linear independent non-orthogonal pure spectral states     can  be discriminated, but  only with a nonzero probability of   inconclusive result \cite{Chefles}. This agrees with   path interference in our case also   between different groups. Only when the spectral states of different groups become orthogonal the cross-group coherence  disappears.

The above conclusions on path interference can be seen directly  from $J$-matrix (which is also unique for a given set of spectral states in contrast to basis-dependent $S$-matrix). Indeed, let us take the  $Q\ge 2$   groups of photons as in the above example. Since permutations of photons in each group between themselves do not change the spectral states, the corresponding $J$-matrix (\ref{E30}) factorizes into a tensor product. Indeed,  let us decompose a permutation $\sigma = \tau\pi$, where  $\tau$ exchanges photons between different groups (without exchanging the order within each group) and $\pi$ exchanges photons within each group. We then have a property $J(\sigma_1,\sigma_2) = J_R(\tau_1,\tau_2)$, which in matrix form reads (compare with Eq. (\ref{E18}) and (\ref{E20})) 
\be
J = J_R\otimes \left(\sum_{q=1}^Q {\!}^{\oplus} v^\dag_q v_q\right), \; J_R(\tau_1,\tau_2) = \langle\vec{\varphi}|P_{\tau_1}P^\dag_{\tau_2}|\vec{\varphi}\rangle,
\en{JR} 
where $v_q $ is defined in Eq. (\ref{S_gen}), $|\vec{\varphi}\rangle = \prod_{q=1}^Q{\!}^{\otimes} \left(|\varphi_q\rangle^{\otimes c_q}\right)$, and the  reduced $J_R$-matrix accounts for interference between photons from different groups ($J(\sigma_1,\sigma_2)$ with the above property  is indeed  a matrix tensor product: if $C= A\otimes B$  the double index notation reads  $C_{ij,kl} = A_{ik}B_{jl}$, in our case $\sigma_i = \tau_i\pi_i$, $i=1,2$ with $\tau_{1,2}$  being the indices of $J_R$ and $\pi_{1,2}$ the indices of $\sum_{q} {\!}^{\oplus} v^\dag_q v_q$). Observing that  summation  over in-group permutations $\pi$ in the product $\prod_{\alpha=1}^NU_{k_{\pi(\alpha)},l_\alpha}$  of Eq.  (\ref{E5}) gives the product of $Q$ quantum amplitudes, one  from each group of photons, we can pass directly to  the argument below Eq. (\ref{P_Snew}) now generalized to $Q$ groups of photons.

\subsubsection{General case: Zero probability conjecture}
\label{sub2}

Now let us consider a general (single photon per mode)  input and  non-ideal (generally dissimilar) detectors. It is clear that non-ideal detectors can result in an effective linear dependence of  spectral states of photons that are otherwise  linearly  independent.  Consider the above  example of $Q$ groups of photons, with $c_q$ photons in the $q$th group having a spectral  state $|\varphi_q\rangle$. For  non-ideal detectors,  if permuted vectors $P_\tau[ \prod_{q=1}^Q{\!}^{\otimes} \left(|\varphi_q\rangle^{\otimes c_q}\right) ]$ for different $\tau$ (permuting vectors between the groups without changing  order within each group) are still linearly independent now under the generalized inner product in  $\mathcal{H}^{\otimes N}$ with the kernel $\hat{K}_{\vec{l}}\equiv \prod_{\alpha=1}^N{\!}^{\otimes} \hat{\Gamma}_{l_\alpha}$,  the above consideration still applies, with the same conclusion about the zero output probability.   The above condition is equivalent to  $\mathrm{det}(\mathcal{G}^{(\alpha)})\ne0 $ for  all $\alpha=1,\ldots,N$, where 
$\mathcal{G}^{(\alpha)}_{ij} = \langle\varphi_i|\hat{\Gamma}_{l_\alpha}|\varphi_j\rangle$. 

Form the above  consideration it is clear that though general detectors   modify linear dependence of spectral states, they  still  can be effectively accounted for (after scaling out their effect on the detection probability, as in section \ref{sec2C}) by considering  another input case with different linear dependence  properties of  spectral states of photons.     Can an output  probability for  only  partially indistinguishable photons vanish exactly  for more  general linear dependent spectral states? In   Ref. \cite{SU3}, where a three-photon coincidence probability was analyzed,  it was found  that dissimilar  detectors strongly influence the coincidence probability  for single photons: it can  be numerically   close to zero  for a non-zero difference of  photon arrival times, if sensitivities of  detectors are strongly different.    However, this cannot be an exact zero probability.  Indeed, in the example  considered in section \ref{sub1}   an exact cancellation  is possible (for a special network) and, by the above  change of  kernel in an inner product, now is  extended to   detector sensitivities resulting in a non-singular kernel,  but   the relevant  condition is still  formulated for \textit{completely indistinguishable} photons (e.g., does not depend on non-zero time delays).  In a more general  case, when detectors result in a singular kernel, this is still true. Let us analyze the  example of three photons with only two linearly independent spectral states.  Indeed, in this case we have $|\varphi_3\rangle = c_1|\varphi_{1}\rangle +c_2|\varphi_{2}\rangle$ for some $c_{1,2}$ and linearly independent $|\varphi_{1,2}\rangle$. We will employ the $S$-matrix approach with the dual basis $\langle j|$, $j=1,2$. In this case there are two sets of $S$-matrices contributing to output probabilities. They correspond to two choices of  three indices $ (j_1,j_2,j_3)$: (i)  $(1,2,1)$ and  permutations $\tau\in \{ I, (1,2), (2,3)\} $ of this set;  or (ii) $(1,2,2)$ and  permutations   $\{ I, (1,2), (1,3)\} $ of this set.  The respective $S$-matrices read (compare with Eq. (\ref{Snew}))
\be
S^{(i)} = \mathcal{M}(\tau)\left(\begin{array}{ccc} 
1& 0& c_1\\
 0&1& c_2\\ 
 1& 0 & c_1\end{array}\right),\;
 S^{(ii)} = \mathcal{M}(\tau)\left(\begin{array}{ccc} 
1& 0& c_1\\
 0&1& c_2\\ 
 0& 1 & c_2\end{array}\right).
 \en{S3}
In the two cases an exact zero  output probability corresponds to a set of equations for the respective quantum amplitudes. Dividing the amplitudes of case $(i)$ of Eq. (\ref{S3}) by $c_i$, $i=1,2$ (thus we assume $c_i\ne0$ otherwise we are in the already  considered case) and setting $\hat{U}_{\alpha,\beta}\equiv U_{k_\alpha,l_\beta}$ we obtain the two sets as follows. For $\tau = I, (1,2), (2,3) $  in Eq. (\ref{S3})  set  (i) reads  
\begin{eqnarray}
\label{set1}
\hat{U}_{11}\hat{U}_{22}\hat{U}_{33} + \hat{U}_{13}\hat{U}_{22}\hat{U}_{31} &=&0,\nonumber\\
\hat{U}_{21}\hat{U}_{12}\hat{U}_{33} + \hat{U}_{23}\hat{U}_{12}\hat{U}_{31} &=&0,\nonumber\\
\hat{U}_{11}\hat{U}_{32}\hat{U}_{23} + \hat{U}_{13}\hat{U}_{32}\hat{U}_{21} &=&0,
\end{eqnarray}
whereas set (ii) for $\tau = I, (1,2), (1,3) $  in Eq. (\ref{S3})  reads
 \begin{eqnarray}
 \label{set2}
\hat{U}_{11}\hat{U}_{22}\hat{U}_{33} + \hat{U}_{11}\hat{U}_{23}\hat{U}_{32} &=&0,\nonumber\\
\hat{U}_{21}\hat{U}_{12}\hat{U}_{33} + \hat{U}_{21}\hat{U}_{13}\hat{U}_{32} &=&0,\nonumber\\
\hat{U}_{31}\hat{U}_{22}\hat{U}_{13} + \hat{U}_{31}\hat{U}_{23}\hat{U}_{12} &=&0.
\end{eqnarray}
(In each set the second and third equation is obtained by transposition of row indices, as dictated by $\tau$, of the first equation.)  There are six different terms  in  Eqs. (\ref{set1})-(\ref{set2}), each being a   product of  three different single-particle amplitudes. Moreover $\hat{U}_{ii}\ne 0$ for $i=1,2,3$, otherwise  $\hat{U}=0$. We simplify Eqs. (\ref{set1})-(\ref{set2}) by introducing $\gamma_{ij} \equiv \hat{U}_{ij}/\hat{U}_{ii}$ and dividing all equations 
by $\hat{U}_{11}\hat{U}_{22}\hat{U}_{33}$.  From the first equation in each system   we get
\be 
\gamma_{12}\gamma_{23}\gamma_{31} = 1,\quad \gamma_{13}\gamma_{21} \gamma_{32}= 1.
\en{g3}
but the second and third equations  in each system result in  
\be 
\gamma_{12}\gamma_{21} = -1,\; \gamma_{23}\gamma_{32} = -1,\; \gamma_{13}\gamma_{31} = -1.
\en{g2}
Eqs. (\ref{g2}) and (\ref{g3}) are  obviously incompatible (as seen by multiplying  them  in each case).

The above  analysis reveals  that in the examples involving  dissimilar detectors in Ref. \cite{SU3} there is only a nearly zero output probability, since  it occurs for a certain set of  non-zero time delays, and  thus cannot be a generalization of the HOM effect \cite{HOM}. What happens is that strongly dissimilar detectors significantly decrease the probability of detection, as discussed in section \ref{sec2B}.   For reference, in appendix  \ref{appC} we  also  consider output probability  for   Gaussian spectral states of photons  in  the  $({s},{\omega})$-basis. Generalizing, let us  formulate  the following  zero probability conjecture  for  an arbitrary  multi-photon input $\vec{n}$ with   mixed spectral states of photons.

\textbf{Zero probability.}--  \textit{The condition for  exactly zero output probability  of  some output configuration   is an  exact cancellation of path amplitudes    of completely indistinguishable photons (generally, a subset of all input photons). Moreover, in such cases the output probability remains equal to  zero  when    degree of distinguishability (for instance, difference in the arrival times)  between only partially indistinguishable photons  is changed.}

By the above,   zero output probability   generally corresponds to various continuously varying degrees of indistinguishability for $N>2$, as was first established in Ref. \cite{Tichy2014} and  generalized above  to  groups of completely indistinguishable photons.   In case of  two photons there is no possibility of exact cancellation of output amplitude if the photons are not completely indistinguishable, which is a restatement of the HOM effect \cite{HOM,Mandel1991}.  In   case of  three photons with linearly dependent  spectral states,  with photons being only partially indistinguishable pairwise,  an exact zero probability is  not possible  at all as shown above.  We conjecture the zero probability result to hold for any input of the type given in Eq.~(\ref{E1}), general detectors,  and  an arbitrary  unitary network.

\subsection{A model  of the  realistic Boson-Sampling device} 
\label{sec3C}

Consider  identical  photon sources and identical detectors (this case was first   considered   in Ref. \cite{NDBS}).  This is a basic model of  input for an   optical realization of the Boson-Sampling computer \cite{AA} which requires single photons to be as indistinguishable as possible. Single photons from  realistic sources \cite{SPS}, as well as  realistic detectors,  have fluctuating parameters  which  cannot be  compensated for (a postselection is the only way to deal with such  fluctuations at the expense of  increasing the   number of runs of the Boson-Sampling device,  which  decreases  its advantage over classical computers).    Note that, in contrast, any bias between  sources or between detectors can be detected and thus corrected for, without resorting to  the  postselection in a Boson-Sampling experiment. Hence, we    assume that the main error of the realistic  Boson-Sampling  device comes from  fluctuations due to mixed spectral states of photons and unstable detector sensitivities, neglecting  any bias error. We focus on the original proposal of Ref.~\cite{AA}, though it is easy to generalize the results to the Boson-Sampling with variable input \cite{GBS} or to another proposal with  time-bin modes replacing  spatial modes   \cite{TBS} (in this case  spatial indices are  replaced with  time-bin indices). 

One can incorporate  fluctuating   sensitivities of unstable detectors  into   spectral states of photons (see below) or, alternatively, use the generalizer kernel for inner product in $\mathcal{H}^{\otimes N}$ and reduced $\hat{J}$-matrix as discussed in section \ref{sec2C}. Consider the corresponding partial indistinguishability matrix $J$. From Eq. (\ref{E48})  we obtain 
\begin{eqnarray}
\label{E51}
J(\sigma_1,\sigma_2) &= &\int\rd x_1p(x_1)\cdot\ldots\cdot\int\rd x_N p(x_N)  \nonumber\\
& \times& \prod_{\alpha=1}^N   \langle\phi(x_{\sigma_1(\alpha)})|\hat{\Gamma}|\phi(x_{\sigma_2(\alpha)})\rangle.
\end{eqnarray}
The crucial point  (see also Ref. \cite{NDBS}) is that  matrix  element $J(\sigma_1,\sigma_2)$ of Eq. (\ref{E51}) actually depends only on the cycle structure of the relative permutation $\sigma_R \equiv \sigma_2\sigma^{-1}_1$, where the cycle structure is $(C_1,\ldots,C_N)$ with $C_k$ being the number  of occurrences in the cycle decomposition of  a cycle  of length $k$  \cite{Stanley}. 
Indeed, due to identical detectors, $J(\sigma_1,\sigma_2)$ of Eq. (\ref{E51})   depends only on  cycle decomposition of the relative permutation $\sigma_R$, as is shown in section \ref{sec2} (see Eq. (\ref{JFACT})).  The  cycle decomposition  factorizes the product  $\prod_{\alpha=1}^N   \langle\phi(x_\alpha)|\hat{\Gamma}|\phi(x_{\sigma_R(\alpha)})\rangle$ into similar products for each cycle. Thanks to the same    probability function $p(x)$ for all single photons the indices of  integration variables $x_\alpha$  are not important, thus     two cycles of the same length (number of elements) contribute the same factor.  Each  factor corresponding to a $k$-cycle of the relative permutation (equivalent to $x_j\to x_{j+1}$, for $j=1,\ldots,k$ with $k+1=1$, by some relabeling of the integration variables) can be  cast  as   follows
\begin{eqnarray*}
&&\int\rd x_1p(x_1)\cdot\ldots\cdot\int\rd x_k p(x_k)\prod_{j=1}^k \langle\phi(x_j)|\hat{\Gamma}|\phi(x_{j+1})\rangle  \\
&& = \mathrm{Tr}\left\{\left(\sqrt{\hat{\Gamma}}\hat{\rho}\sqrt{\hat{\Gamma}}\right)^k \right\}.
\end{eqnarray*}
Therefore, we get the following formula for the partial indistinguishability matrix 
\be
J(\sigma_1,\sigma_2)  = \prod_{k=1}^N g_k^{C_k(\sigma_2\sigma^{-1}_1)},\quad g_k \equiv \mathrm{Tr}\left\{\left(\sqrt{\hat{\Gamma}}\hat{\rho}\sqrt{\hat{\Gamma}}\right)^k \right\}. 
\en{E52}
It is easy to see from the definition that  parameters $0\le g_k\le 1$, describing  partial indistinguishability of  single  photons from identical sources,  satisfy the constraint $g_{k+m} \le g_kg_m$ which  indicates that generally one will have  decrease of  indistinguishability of photons with increase of the number of sources (see also Fig. \ref{F2} below).   

Eq. (\ref{E52})  implies that   detector sensitivities  can be dealt with by  introducing  an  (unnormalized) spectral state of  a photon  visible to a detector  as follows
\be
\Phi(s,\omega;x,y) \equiv \phi(s,\omega;x)\sqrt{\Gamma(s,\omega;y)},
\en{Phi_new}
where $y$ is some fluctuating parameter(s) of the detector. One can easily see that in this case the corresponding reduced  $\hat{J}$-matrix is given as $J$-matrix (\ref{E51}) with $\hat{\Gamma}=I$ and spectral states of Eq.  (\ref{Phi_new}).

Let us consider in some detail the case of   single photons with a fixed polarization and  random arrival times, when   their spectral function (augmented by detector sensitivities)  is a   Gaussian 
 \be
\Phi(\omega;\tau) = \left(2\pi\Delta\omega^2\right)^{-\frac14}\exp\left( i\omega\tau - \frac{\omega^2}{4\Delta\omega^2}\right),
\en{E53}
as well as the distribution of their  arrival times
\be
p(\tau) = \frac{1}{\sqrt{2\pi}\Delta\tau}\exp\left(   - \frac{\tau^2}{2\Delta\tau^2}\right).
\en{E54}
 We have  (see   also Ref. \cite{NDBS}) $g_k = (1-\gamma)^{k/2}(1-\gamma^k)^{-1/2}$ where $\gamma = 2\eta^2/(1+2\eta^2)$  with $\eta = \Delta\omega\Delta\tau$ being the classicality parameter (the case of  completely indistinguishable photons  corresponds to  $\eta=0$, whereas for maximally distinguishable photons $\eta=\infty$).  The partial indistinguishability matrix reads  \cite{NDBS}
\be
J(\sigma_1,\sigma_2) = (1-\gamma)^\frac{N}{2}\prod_{k=1}^N (1-\gamma^k)^{-\frac{C_k}{2}}, 
\en{E55}
where $(C_1,\ldots,C_N)$ is the cycle structure of $\sigma_2\sigma^{-1}_1$. 

To measure how close is the matrix $J$ of Eq. (\ref{E55})   to the case of completely indistinguishable photons, let us study its purity  defined  in Eq. (\ref{Purity}) of section \ref{subPurity}. 
We have 
\begin{eqnarray}
\label{E57}
&& \mathrm{Tr}\left\{\left(\frac{J}{N!}\right)^2\right\} = \frac{(1-\gamma)^N}{N!}\sum_\sigma\prod_{k=1}^N (1-\gamma^k)^{-C_k(\sigma)} \nonumber\\
&& = (1-\gamma)^N Z_N(1/(1-\gamma),\ldots,1/(1-\gamma^N)),
\end{eqnarray}
where $Z_N=Z_N(a_1,\ldots,a_N)$, the divided by $N!$ sum of  powers $\prod_{k=1}a_k^{C_k}$ over all permutations,  is known as the cycle index  for which there is  a generating function \cite{Stanley}
\be
F(x) \equiv \sum_{N\ge 0} Z_N(a_1,\ldots,a_N)x^N = \exp\left( \sum_{k=1}^\infty \frac{a_kx^k}{k}\right).
\en{E56}
In our case $a_k = 1/(1-\gamma^k)$ and we obtain
\be
\sum_{k=1}^\infty \frac{x^k}{k(1-\gamma^k)} = \sum_{l=0}^{\infty}\sum_{k=1}^{\infty} \frac{(\gamma^lx)^k}{k} = -\sum_{l=0}^\infty\ln\left(1-\gamma^lx\right).
\en{E59}
Using  the following  identity involving   q-Pochhammer symbol $(x;q)_N  \equiv \prod_{k=0}^{N-1}(1-xq^k)$
\[
\prod_{k\ge0}(1-x\gamma^k) = \sum_{N\ge0} \frac{x^N}{(\gamma;\gamma)_N},
\]
from Eqs. (\ref{E57})-(\ref{E59})  we  obtain 
\be
\mathrm{Tr}\left\{\left(\frac{J}{N!}\right)^2\right\}  = \frac{(1-\gamma)^N}{\prod_{k=1}^N(1-\gamma^k)}.
\en{E58}
Eq. (\ref{E58}) is  the  law of  purity (and, therefore, the   indistinguishability) decrease with increase of the number of  sources $N$ and/or  the  classicality parameter $\gamma$ of each source.   For small $\gamma\ll1$ (i.e.,  $\eta^2\ll1$) we obtain $\mathrm{Tr}\left\{\left(\frac{J}{N!}\right)^2\right\}  \approx 1 - 2(N-1)\eta^2$. The behavior of $\mathcal{P}$ with $\gamma$ for various $N$ is illustrated in Fig.~\ref{F2}. 

Finally,   small  bias errors can be  considered similarly as in  Ref. \cite{BSscal}.   

\begin{figure}[htb]
\begin{center}
\includegraphics[width=0.45\textwidth]{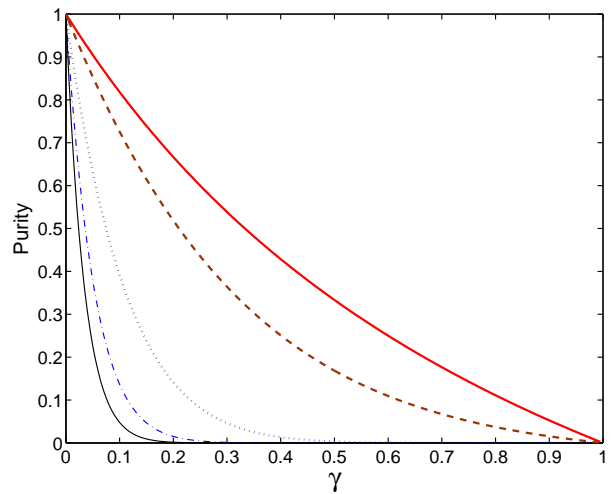}
\caption{ (Color online)  Purity $\mathcal{P}$ (\ref{Purity}) of the partial indistinguishability matrix $J$ vs. the parameter $\gamma$. Here: $N=2$ (thick solid line), $N=4$ (dashed line), $N = 10$ (dotted line), $N = 20$ (dash-dotted line), and $N=30$ (thin solid line).    }\label{F2}
\end{center}
\end{figure}

\section{Conclusion}
\label{con}

We have developed  a theory of    partial indistinguishability  of photons for   multi-photon experiments in multi-port devices.  The key object is the partial indistinguishability matrix, a  non-negative definite Hermitian matrix built from spectral states of   photons and detector sensitivities. Though only a fraction of information in  the partial indistinguishability matrix seems to be derivable from the corresponding  output probabilities,  using an expression for output probability as  a quadratic form  and a  clear physical interpretation of its  arguments as path amplitudes is quite appealing, moreover, it  allows   physical insights. For instance, a connection with complementarity of the which-way vs.  the interference visibility is used in   formulation of the  zero probability conjecture.
 The permutation (symmetric) group is the key object of the theory,  the partial indistinguishability matrix is indexed by permutations of  photonic  spectral states and has the dimension $N!\times N!$ for $N$ photons.  It  is interesting to note that  the advanced features of the group,  such as   non-trivial group characters  and the matrix immanants,  related to them,   do not play any  role in our approach. For instance, we have shown that  output probability is always expressed in terms of the matrix permanents only (the matrix permanent is  related to  the trivial character of the permutation group).    In special cases  the partial indistinguishability matrix reduces to  much simpler  forms, amenable for even an analytical analysis.    We have also found that a possible generalization of  Mandel's indistinguishability parameter for $N>2$ photons is  given by the purity of a reduced partial indistinguishability matrix, where only the effect of detectors on partial indistinguishability is retained, whereas their effect on the total probability is scaled out.  We have found an analytical expression  giving  the  purity measure of quantum coherence  for  a model of a realistic Boson-Sampling computer. Besides  experiments with optical multiports, the theory can be applied also to   quantum walks with several photons \cite{QW1,QW2,QW3} where indistinguishability of  photons  is  essential for  such multi-particle walks to show  quantum  correlations of a   many-boson system. The approach  developed here  was already used for derivation of    very interesting  results in Ref. \cite{Tichy2014}.

\section{Acknowledgements}

This work was supported by the CNPq (Brazil).  A part of this work was done during the   visit to B.I.~Stepanov  Institute of Physics,    National Academy of Sciences of Belarus. The author is grateful to Dmitri Mogilevtsev for many invaluable discussions. The author is indebted to Malte C. Tichy for  pointing out a loophole in the initial  formulation of the zero output probability result.

\appendix
\section{Derivation of the probability formula}
\label{appA}

We will use the following identity

\begin{eqnarray}
&&\langle0|\left[\prod_{\alpha=1}^Nb_{l_\alpha,s_\alpha}(\omega_\alpha)\right] \left[\prod_{\alpha=1}^Nb^\d_{l^\pr_\alpha,s^\pr_\alpha}(\omega^\pr_\alpha)\right]|0\rangle \nonumber\\
&&= \sum_{\sigma}\prod_{\alpha=1}^N\delta_{l^\pr_\alpha,l_{\sigma(\alpha)}} \delta_{s^\pr_\alpha,s_{\sigma(\alpha)}} \delta(\omega^\pr_\alpha-\omega_{\sigma(\alpha)}),
\label{A1}\end{eqnarray}
where the summation is over all permutations $\sigma$ of $N$ elements.  Inserting Eqs. (\ref{E1}) and (\ref{E3}) into Eq. (\ref{E4})  we obtain
\begin{eqnarray}
\label{A2}
&& P(\vec{m}|\vec{n}) =  \frac{1}{\mu(\vec{m})\mu(\vec{n})} \sum_{\vec{s}} \sum_{\vec{s}{\,}^\pr}\sum_{\vec{s}^{\pr\prime}} \int  \rd \vec{\omega}  \int\rd\vec{\omega}^\pr\int\rd\vec{\omega}^{\pr\pr} \nonumber\\
&&\times \left[\prod_{\alpha=1}^N\Gamma_{l_\alpha}(s_\alpha,\omega_\alpha)\right] G(\vec{s}^{\,\pr},\vec{\omega}^\pr|\vec{s}^{\,\pr\pr},\vec{\omega}^{\pr\pr})\nonumber\\
&& \times \langle0|  \left[\prod_{\alpha=1}^N a_{k_\alpha,s^{\pr\pr}_\alpha}(\omega_\alpha^{\pr\pr}) \right]\left[\prod_{\alpha=1}^Nb^\d_{l_\alpha,s_\alpha}(\omega_\alpha)\right]|0\rangle \nonumber\\
&& \times \langle0| \left[\prod_{\alpha=1}^Nb_{l_\alpha,s_\alpha}(\omega_\alpha)\right]  \left[\prod_{\alpha=1}^N a^\d_{k_\alpha,s^\pr_\alpha}(\omega_\alpha^\pr)\right]|0\rangle.
\end{eqnarray}
 By   using the network transformation  $a^\d_{k,s}(\omega) = \sum_{l=1}^M U_{kl}b^\d_{l,s}(\omega)$ and Eq. (\ref{A1}) we get, for instance,
 \begin{eqnarray*}
% \label{A3}
&&   \langle0|  \left[\prod_{\alpha=1}^N a_{k_\alpha,s^{\pr\pr}_\alpha}(\omega_\alpha^{\pr\pr}) \right]\left[\prod_{\alpha=1}^Nb^\d_{l_\alpha,s_\alpha}(\omega_\alpha)\right]|0\rangle \nonumber\\
 &&= \sum_{\vec{l}^{\pr\pr}}\left[ \prod_{\alpha=1}^N U^*_{k_\alpha,l^{\pr\pr}_\alpha} \right]\sum_\sigma \prod_{\alpha=1}^N\delta_{l^{\pr\pr}_\alpha,l_{\sigma(\alpha)}} \delta_{s^{\pr\pr}_\alpha,s_{\sigma(\alpha)}} \delta(\omega^{\pr\pr}_\alpha-\omega_{\sigma(\alpha)})\nonumber\\
 && = \sum_\sigma \left[ \prod_{\alpha=1}^N U^*_{k_\alpha,l_{\sigma^{-1}(\alpha)}} \right]  \prod_{\alpha=1}^N  \delta_{s^{\pr\pr}_\alpha,s_{\sigma(\alpha)}} \delta(\omega^{\pr\pr}_\alpha-\omega_{\sigma(\alpha)}).
 \end{eqnarray*}
This identity  and a similar relation for the second inner product in Eq. (\ref{A2}) transform Eq. (\ref{A2}) to a resulting expression equivalent to Eq. (\ref{E5}) of section \ref{sec2}. The final step is to transfer  permutations from the $l$-indices to the $k$-indices in the two products of  network matrix elements  by using the following general  identity for any two permutations  $\sigma$ and $\tau$
\be
\prod_{\alpha} A_{\alpha,\tau\sigma(\alpha)} = \prod_{\alpha} A_{\sigma^{-1}(\alpha),\tau(\alpha)},
\en{A3}
which easily follows from independence of a product of scalars from their order and the fact that a permutation is just a bijection between two sets of indices.

\section{$S$-matrices  not contributing to output probability }
\label{appB}

 Consider $Q\ge 2$   groups of photons, where group $q$  consists of photons in a spectral state $|\varphi_q\rangle$,  the spectral states $|\varphi_1\rangle,\ldots,|\varphi_Q\rangle$ being  linearly independent.  What  choice of $\vec{j}$ in matrix $S(\vec{j})$ trivially  results in zero output  probability in Eq. (\ref{P_new}) (i.e., irrespective $U$)?  Let  $c_q$ be  the number of photons in the spectral state $|\varphi_q\rangle$.   If $|\vec{j}\rangle $ is a tensor product of vectors which do not represent a permutation of the dual basis set $|1\rangle\otimes\ldots\otimes|1\rangle\otimes|2\rangle$ then the corresponding matrix  $S(\vec{j})$ consists of non-square (rectangular) blocks  of entries equal to $1$, whereas  the complementary blocks have zeros in each entry. Then, irrespective of  a network matrix $U$, the matrix permanent of the Hadamard product of    matrices  $S$ and $U[\vec{n}|\vec{m}]$ can be expanded by using the analog of Laplace formula for a permanent  of an $N\times N$-dimensional matrix \cite{Minc}
 \begin{eqnarray}
 \label{B1}
 \mathrm{per}(A) &=& \sum_{1\le i_1<...<i_k\le N}  \mathrm{per}(A[1,...,k|i_1,...,i_k]) \nonumber\\
 &\times& \mathrm{per}(A[k+1,...,N|i_{k+1},...,i_N]),
 \end{eqnarray}
where $(i_1,...,i_N)$ is a permutation of $(1,..,,N)$ and we have divided matrix $A$ into two square blocks of dimension $k$ and $N-k$. Now, by the structure of matrix $S(\vec{j})$ for $\vec{j}$ not a permutation of the dual basis,  the permanent of one of the blocks of $S$  in each term in a sum similar to that of Eq. (\ref{B1}) is always equal to zero, since there are sets of $k$ rows (or columns) of such a matrix $S$   containing \textit{strictly less}  then $k$ columns (rows) which are nonzero.

\section{Photons in pure Gaussian states}
\label{appC}
 
For strongly dissimilar detectors  output probabilities approach zero  (for some or even all output configurations),  if the product of detector sensitivities  approaches zero, simply   due to the fact  that there are also  detection probabilities in this case, i.e., given by  matrix  $D_{\vec{m}}$ of section \ref{sec2C}.   Following 
Ref. \cite{SU3}, let us consider single  photons  of the same  polarization  and with  Gaussian spectral functions  of  center frequencies $\Omega_\alpha$  and  arrival times $t_\alpha$. Thus 
\be
\phi_\alpha(\omega) = (2\pi\epsilon\Delta_\alpha^2)^{-\frac14}\exp\left\{it_\alpha\omega-\frac{(\omega-\Omega_\alpha)^2}{4\epsilon\Delta_\alpha^2}\right\},
\en{E37} 
where   we have inserted   $\epsilon>0$  to study  the limit of monochromatic photons   (see below).  We have from Eq. (\ref{E30})  of section \ref{sec3A}
\begin{eqnarray}
\label{E39}  
&& J(\sigma_1,\sigma_2) = \int\rd\vec{\omega}\, \left[\prod_{\alpha=1}^N(2\pi\epsilon\Delta^2_{\alpha})^{-\frac12}\Gamma_{l_\alpha}(\omega_\alpha)\right]\nonumber\\
&&\times \exp\biggl\{ -\sum_{\alpha=1}^N \sum_{i=1,2}\frac{(\omega_\alpha-\Omega_{{\sigma_i(\alpha)} })^2}{ 4\epsilon\Delta^2_{{\sigma_i(\alpha)}} }  \nonumber\\
&&  +i\sum_{\alpha=1}^N\omega_\alpha(t_{{\sigma_2(\alpha)} }- t_{{\sigma_1(\alpha)}}) \biggr\}.
\end{eqnarray}
Let us consider output  probability   for  $J(\sigma_1,\sigma_2)$ of Eq.~(\ref{E39}),  for arbitrary    detector sensitivities. Indeed, output probability  in this case can be easily rewritten as follows (setting $\epsilon =1 $)
\begin{eqnarray}
\label{PGauss}
&& P(\vec{m}) = \int\rd\vec{\omega}\,\Bigl|\sum_{\sigma}Z_\sigma(\vec{\omega})\Bigr|^2,\nonumber\\
&&  Z_\sigma(\vec{\omega}) \equiv  \prod_{\alpha=1}^N (2\pi\Delta_{\alpha}^2)^{-\frac14}\sqrt{\Gamma_{l_\alpha}(\omega_\alpha)}X_{\sigma(\alpha)}(\omega_\alpha),\qquad 
\end{eqnarray}
where $X_\beta(\omega_\alpha) =  \exp\left\{i\omega_\alpha t_{\beta}-\frac{(\omega_\alpha -\Omega_{\beta})^2}{4\Delta_{\beta}^2}\right\}U_{k_{\beta},l_\alpha}$. 
The sum in Eq. (\ref{PGauss}) is nothing but the matrix permanent, we have
\begin{eqnarray}
&& P(\vec{m}) = \int\rd\vec{\omega}\, |\mathrm{per}\bigl(V(\vec{\omega})\bigr)|^2,\nonumber\\
&& V_{\beta,\alpha}(\vec{\omega})\equiv (2\pi\Delta_{\alpha}^2)^{-\frac14}\sqrt{\Gamma_{l_\alpha}(\omega_\alpha)} X_{\beta}(\omega_\alpha).
\label{perZ}
\end{eqnarray}
For $P(\vec{m}) $   of Eq. (\ref{PGauss}) to be zero  requires that $\sum_\sigma Z_\sigma(\vec{\omega}) =0$ is zero  at any point $\vec{\omega}$.     We note that  sum $\sum_\sigma Z_\sigma(\vec{\omega})$ can   be rather  close to zero,   when detectors have strongly dissimilar sensitivities. Precisely this happens in the examples of Ref. \cite{SU3}).

In the limit of monochromatic    single photons, $\epsilon\to0$,  the above expressions reduce to those of  the two extreme cases discussed in section \ref{sec2B}.     In this limit one  does not need to specify  detector sensitivities  as only some  point  values will be needed.  Using the following expansion in powers of $\epsilon$ 
\begin{eqnarray}
\label{E40}
&& \frac{1}{\sqrt{2\pi}\epsilon}\exp\left\{-\frac{1}{2\epsilon^2}\sum_{i=1,2}\frac{(\omega-\Omega_i)^2}{2\Delta_i^2} \right\} = \delta_{\Omega_1,\Omega_2}
 \delta(\omega-\Omega_1)\nonumber\\
&& \times \frac{\sqrt{2}\Delta_1\Delta_2}{\sqrt{\Delta^2_1+\Delta^2_2}}  + O(\epsilon),
\end{eqnarray}
we easily obtain from Eq. (\ref{E39})
\begin{eqnarray}
\label{E41}
&&J(\sigma_1,\sigma_2) = F(\sigma_1,\sigma_2)\left[\prod_{\alpha=1}^N \delta_{\Omega_{{\sigma_1(\alpha)}},\Omega_{{\sigma_2(\alpha)}}}
 \right]\nonumber\\
&& \times \exp\left\{i\sum_{\alpha=1}^N \Omega_{{\sigma_1(\alpha)}}\left(t_{{\sigma_2(\alpha)}}-t_{{\sigma_1(\alpha)}}\right)\right\} + O(\epsilon), \qquad\end{eqnarray}
where we have  set
\be
F(\sigma_1,\sigma_2) \equiv \prod_{\alpha=1}^N \Gamma_{l_\alpha}(\Omega_{{\sigma_1(\alpha)}})\left[\frac{2\Delta_{\sigma_1(\alpha)}\Delta_{\sigma_2(\alpha)}}{\Delta^2_{\sigma_1(\alpha)}+\Delta^2_{\sigma_2(\alpha)}}\right]^\frac12. 
\en{E42}
It immediately follows that if frequencies $\Omega_\alpha$ of  monochromatic single photons are pairwise different then the corresponding partial indistinguishability matrix $J$   (\ref{E41})  is diagonal (i.e., maximally mixed) $J(\sigma_1,\sigma_2) = D(\sigma_1)\delta_{\sigma_1,\sigma_2}$ with 
\be
D(\sigma_1)  \equiv \prod_{\alpha=1}^N\Gamma_{l_\alpha}(\Omega_{\sigma_1(\alpha)}). 
\en{E43}
  (compare with Eq. (\ref{E26}) of section \ref{sec2B}).  In this case  monochromatic photons behave in a way similar  to that of  classical particles. 
  
   In the opposite extreme case, when single photons have equal frequencies, $\Omega_\alpha=\Omega$, assuming also the same spectral width, $\Delta_\alpha = \Delta$,   we get from Eq. (\ref{E41})  $J(\sigma_1,\sigma_2) = D$, 
   where $D$ is  of  Eq. (\ref{E43}) with   $\Omega_\alpha= \Omega$ (compare with Eq. (\ref{E25}) of section \ref{sec2B}). Output  probability in this case is the same as for completely  indistinguishable photons, i.e., 
\be
P(\vec{m}) = \frac{D}{\mu(\vec{m})} |\mathrm{per}(U[\vec{n}|\vec{m}])|^2.
\en{E45}

%%%%%%%%%%%%%

\end{document}